\documentclass{aa}
\usepackage{epsfig}
\usepackage{natbib}
\usepackage{graphics}
\usepackage{color} \usepackage{ulem}
\usepackage{color}
\newfont{\tss}{cmssdc10 scaled 950}

\begin{document}

\title{Tol 2240--384 - a new low-metallicity AGN candidate\thanks{Based on 
observations
collected at the European Southern Observatory, Chile, ESO program 
69.C-0203(A), 71.B-0509(A) and 383.B-0271(A).}
% and at the CDS via
%anonymous ftp to cdsarc.u-strasbg.fr (130.79.128.5) or via
%http://cdsweb.u-strasbg.fr/cgi-bin/qcat?J/A+A/?/?}
}
%\subtitle{Discovery of the most metal-poor  H~{\sc ii} regions  in SBS 0335-052W}

\author{Y. I.\ Izotov \inst{1,2,4}
\and N. G.\ Guseva \inst{1,2}
\and K. J.\ Fricke \inst{1,3}
\and G.\ Stasi\'nska \inst{4}
\and C.\ Henkel \inst{1}
\and P.\ Papaderos \inst{5,6}}
\offprints{Y.I. Izotov, izotov@mao.kiev.ua}
\institute{          Max-Planck-Institut f\"ur Radioastronomie, Auf dem H\"ugel 
                     69, 53121 Bonn, Germany
\and
                     Main Astronomical Observatory,
                     Ukrainian National Academy of Sciences,
                     Zabolotnoho 27, Kyiv 03680,  Ukraine
\and
                     Institut f\"ur Astrophysik, 
                     G\"ottingen Universit\"at, Friedrich-Hund-Platz 1, 
                     37077 G\"ottingen, Germany
\and
                     LUTH, Observatoire de Paris, CNRS, Universite Paris Diderot,
                     Place Jules Janssen 92190 Meudon, France
\and
                     Centro de Astrof\'isica da Universidade do Porto, Rua das 
                     Estrelas, 4150-762 Porto, Portugal
\and
                     Department of Astronomy, Oskar Klein Centre, Stockholm 
                     University, SE - 106 91 Stockholm, Sweden
}

\date{Received \hskip 2cm; Accepted}

\abstract
% Context
{ Active galactic nuclei (AGNs) have typically been discovered in 
massive galaxies of high metallicity.} 
% Aims
{We attempt to increase the number of AGN candidates in low metallicity
  galaxies. 
We present VLT/UVES and archival VLT/FORS1 spectroscopic and NTT/SUSI2 photometric observations
of the low-metallicity  emission-line galaxy Tol~2240--384 and perform a
detailed study of its morphology, chemical composition, and emission-line profiles.}  
%Methods
{The profiles of emission lines in the UVES and FORS1 spectra are decomposed
into several components with different kinematical properties 
by performing multicomponent fitting with Gaussians. 
We determine abundances of nitrogen, oxygen, neon, 
sulfur, chlorine, argon, and iron by analyzing the fluxes of narrow components 
of the emission lines using empirical methods. 
We verify with a photoionisation model that the physics of the narrow-line 
component gas is similar to that in common metal-poor galaxies. 
} 
% Results
{Image deconvolution reveals two high-surface brightness regions 
in Tol~2240--384 separated by 2.4 kpc.
% and differing in their luminosity by a factor of $\sim$10. 
The brightest southwestern region is surrounded by intense ionised gas 
emission that strongly affects the observed $B-R$ colour
on a spatial scale of $\sim$5 kpc. 
The profiles of the strong emission lines in the UVES spectrum are asymmetric and all 
these lines apart from H$\alpha$ and H$\beta$ can be fitted by two Gaussians 
of FWHM $\sim$ 75 -- 92 km s$^{-1}$ separated by $\sim$ 80 km s$^{-1}$ 
implying that there are two regions of ionised gas emitting narrow lines. 
The oxygen abundances in both regions are equal within the errors
and in the range 12+log O/H = 7.83 -- 7.89. The shapes of the H$\alpha$ and
H$\beta$ lines are more complex. In particular, the H$\alpha$ emission line
consists of two broad components of FWHM $\sim$ 
700 km s$^{-1}$ and 2300 km s$^{-1}$, in addition to narrow components of two 
regions revealed from profiles of other lines. 
This broad emission in H$\alpha$ and H$\beta$ 
associated with the high-excitation, brighter southwestern H {{\sc ii}} 
region of the galaxy
is also present in the archival low- and medium-resolution VLT/FORS1 spectra. 
The extraordinarily high luminosity of the broad H$\alpha$ line 
of 3 $\times$ 10$^{41}$ erg s$^{-1}$ cannot be accounted 
for by massive stars at different stages of their evolution. 
The broad H$\alpha$ emission persists over a period of 7 years, which 
excludes supernovae as a powering
mechanism of this emission. This emission most likely arises from an
accretion disc around a black hole of mass $\sim$ 10$^7$ $M_\odot$.}
% Conclusions
{}

\titlerunning{Tol 2240--384 - a new low-metallicity AGN candidate}
\authorrunning{Y.I.Izotov et al.}
\maketitle

%\markboth{Y.I.Izotov et al.}{Abundance patterns in the emission-line
%galaxies from the SDSS Early Data Release}

\section{Introduction}

Active galactic
nuclei (AGNs) are understood to be powered by massive black holes at the centers of
galaxies, accreting gas from their surroundings. They are usually
found in massive, bulge-dominated galaxies and their gas metallicities are 
generally high \citep{S98,H02,H09}. However, it remains unclear whether 
AGNs in low-metallicity low-mass galaxies do exist. \citet{Gr06} and \citet{B08}
searched the Sloan Digital Sky Survey (SDSS) spectroscopic galaxy samples 
for low-mass Seyfert 2 galaxies.
In particular, \citet{Gr06} used a sample of
23 000 Seyfert 2 galaxies selected by \citet{K03} and found only 
$\sim$ 40 Seyfert 2 galaxies among them with masses 
lower than  10$^{10}$ $M_\odot$.
They demonstrated, however, that the metallicities of these AGNs are
around solar or slightly subsolar. 
The same high metallicity range is found in the SDSS sample of 
174 low-mass broad-line AGNs of 
\citet{G07}. On the other hand, \citet{I07} and \citet{IT08} demonstrated that 
broad-line AGNs with much lower metallicities probably exist, 
although they occupy a region
in the diagnostic diagram differing from that of more metal-rich AGNs and
are extremely rare. They identified
four of these galaxies, which were found to have 
oxygen abundances 12 + log O/H in the range
7.36 -- 7.99 on the basis of a systematic search for   
extremely metal-deficient emission-line dwarf galaxies in
the SDSS Data Release 5 (DR5) database 
of 675 000 spectra.  
The absolute magnitudes of those four low-metallicity AGNs
are typical of dwarf galaxies, their host galaxies have a compact structure,
and their spectra
resemble those of low-metallicity high-excitation H~{\sc ii} regions.
\citet{I07} found that there is however a striking
difference: the strong permitted emission lines, mainly the H$\alpha$
$\lambda$6563 line, show very prominent broad components
characterised by properties unusual for dwarf galaxies: 1) their H$\alpha$ full
widths at zero intensity (FWZI) vary from 102 to 158 \AA,
corresponding to expansion velocities between 2200 and 3500 km
s$^{-1}$; 2) the broad H$\alpha$ luminosities $L_{br}$ are
extraordinarily large, between  3$\times$10$^{41}$ and
2$\times$10$^{42}$ erg s$^{-1}$.  This is higher than the range 
10$^{37}$--10$^{40}$ erg~s$^{-1}$ found by \citet{I07} for the other
emission-line galaxies (ELGs) with broad-line emission.  The ratio of 
H$\alpha$ flux in the
broad component to that in the narrow component varies from 0.4 to
3.4, compared to 0.01--0.20 for the other galaxies; 3) the Balmer
lines exhibit a very steep decrement, which is indicative of 
collisional excitation and the broad emission originating in very dense gas
($N_e$$>$10$^{7}$ cm$^{-3}$).

To account for the broad-line
emission in these four objects, \citet{I07} considered various physical
mechanisms such as Wolf-Rayet (WR) stars, stellar winds from Ofp or luminous
blue variable stars, single or multiple supernova (SN) remnants 
propagating in the interstellar medium, and SN bubbles.
While these mechanisms may be able to produce $L_{br}$ $\sim$ 10$^{36}$ 
to 10$^{40}$ erg s$^{-1}$, they cannot generate yet higher
luminosities, which are more likely associated with SN shocks or AGNs. 
\citet{I07} considered type IIn SNe because their H$\alpha$ 
luminosities are higher ($\sim$10$^{38}$--10$^{41}$ erg~s$^{-1}$) than those 
of the other SN types and they decrease less rapidly. 
\citet{IT08} found no significant temporal 
evolution of broad H$\alpha$ in all four galaxies over a period of 3--7 
years. Therefore, the IIn SNe mechanism may be excluded, leaving
only the AGN mechanism capable of accounting for the high luminosity of the
broad H$\alpha$ emission. However, we also have difficulty with this mechanism.
In particular, all four galaxies are present in neither the ROSAT catalogue 
of the X-ray sources nor the NVSS catalogue of radio sources. 
High-ionisation emission lines such as He {\sc ii} $\lambda$4686 or 
[Ne {\sc v}] $\lambda$3426 are weak or not detected in optical spectra. 
Based on the observational evidence, \citet{IT08} concluded that all four
studied galaxies most likely belong to the very rare type of low-metallicity 
AGNs in which non-thermal ionising radiation is strongly diluted by the 
radiation of a young massive stellar population.

The fifth galaxy of this type, Tol 2240--384, was first spectroscopically 
studied by \citet{T91} and \citet{M94}. However, in those low-resolution 
spectra, some important emission lines, such as [O~{\sc ii}] $\lambda$3727,
[Ne~{\sc iii}] $\lambda$3868, and H$\alpha$ $\lambda$6563, are missing. 
This, in particular, precludes abundance determination and the
detection of broad hydrogen emission. \citet{K04,K06} studied
Tol 2240--384 spectroscopically, and \citet{K06} 
derived the oxygen abundance of
this galaxy, 12+logO/H = 7.77 $\pm$ 0.08. We note that no broad emission was
reported by \citet{T91}, \citet{M94}, and \citet{K06}.
In this paper, we present 8.2m Very Large Telescope (VLT) spectroscopic 
observations
and 3.5 ESO New Technology Telescope (NTT) photometric observations
of this emission-line galaxy. Its optical spectrum shows the very broad 
components of hydrogen emission lines and is similar to those
found previously by \citet{I07} and \citet{IT08} for the four other galaxies.
We describe observations in Sect.2. The morphology of the galaxy is discussed
in Sect.3 and its location in the emission-line diagnostic diagram is
discussed in Sect.4. Element abundances are derived in Sect.5.
The kinematics of the ionised gas from narrow emission lines is discussed
in Sect.6. We discuss in Sect.7 the properties of the broad emission and 
derive the mass of the central black hole assuming an AGN mechanism for the
origin of the broad line emission. Our conclusions are summarized in Sect.8.

%%%%%%%%%%%%%%%%%%%%%%%%%%%%%%%%%%%%%%%%%%%%%%%%
%    Fig.1  (UVES)
%%%%%%%%%%%%%%%%%%%%%%%%%%%%%%%%%%%%%%%%%%%%%%%%
\begin{figure*}[t]
\hspace*{0.5cm}\psfig{figure=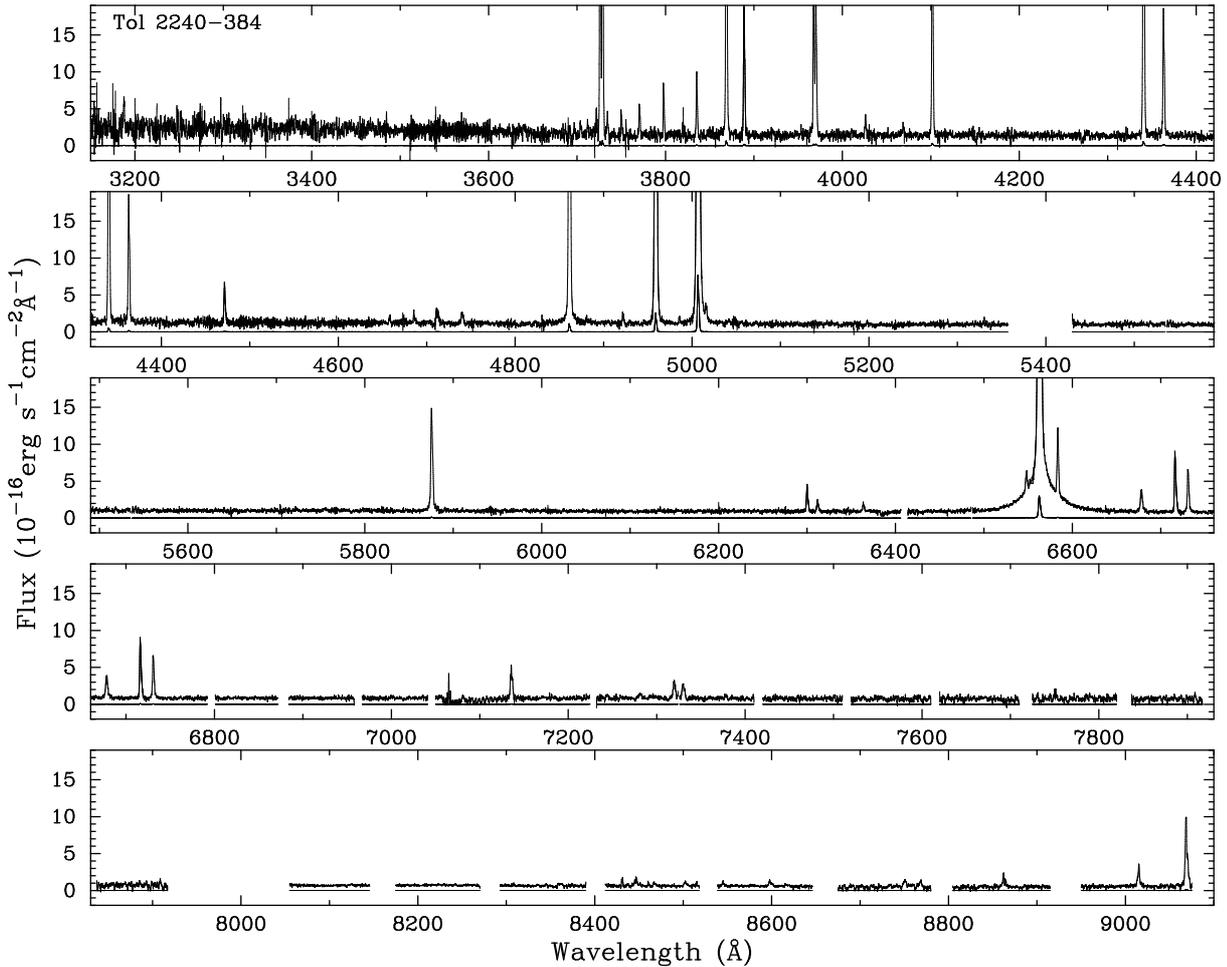,angle=-90,width=16.cm,clip=}
\caption{Flux-calibrated VLT/UVES spectrum 
of Tol 2240--384, obtained on 23 August 2009, corrected for the redshift 
of $z$ = 0.07595 [ESO program 383.B-0271(A)] (upper spectrum).
The lower spectrum is the upper spectrum downscaled by a factor of 100.
The scale of the ordinate is that for the upper spectrum.
Note the broad emission in the hydrogen line H$\alpha$ $\lambda$6563.
No appreciable broad emission is detected in other strong 
permitted and forbidden lines, which is indicative of the rapid
motions of relatively dense ionised gas with an electron number
density $N_e$ $\geq$ 10$^7$ cm$^{-3}$.}
\label{fig1}
\end{figure*}
%%%%%%%%%%%%%%%%%%%%%%%%%%%%%%%%%%%%%%%%%%%%%%%%%

%%%%%%%%%%%%%%%%%%%%%%%%%%%%%%%%%%%%%%%%%%%%%%%%
%    Fig.2 (FORS)
%%%%%%%%%%%%%%%%%%%%%%%%%%%%%%%%%%%%%%%%%%%%%%%%
\begin{figure*}[t]
\hspace*{1.0cm}\psfig{figure=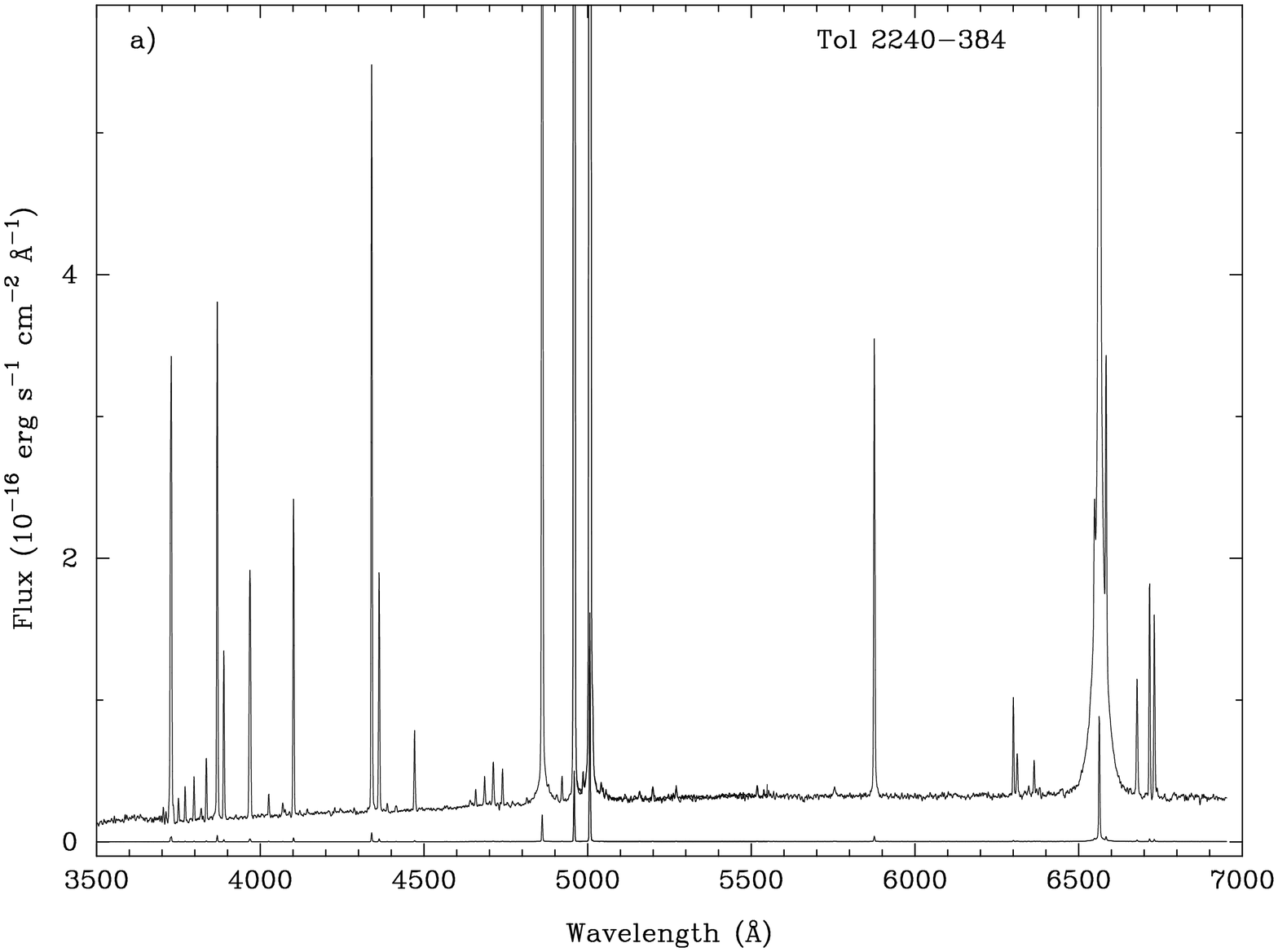,angle=-90,width=7.5cm}%,clip=}
\hspace*{1.0cm}\psfig{figure=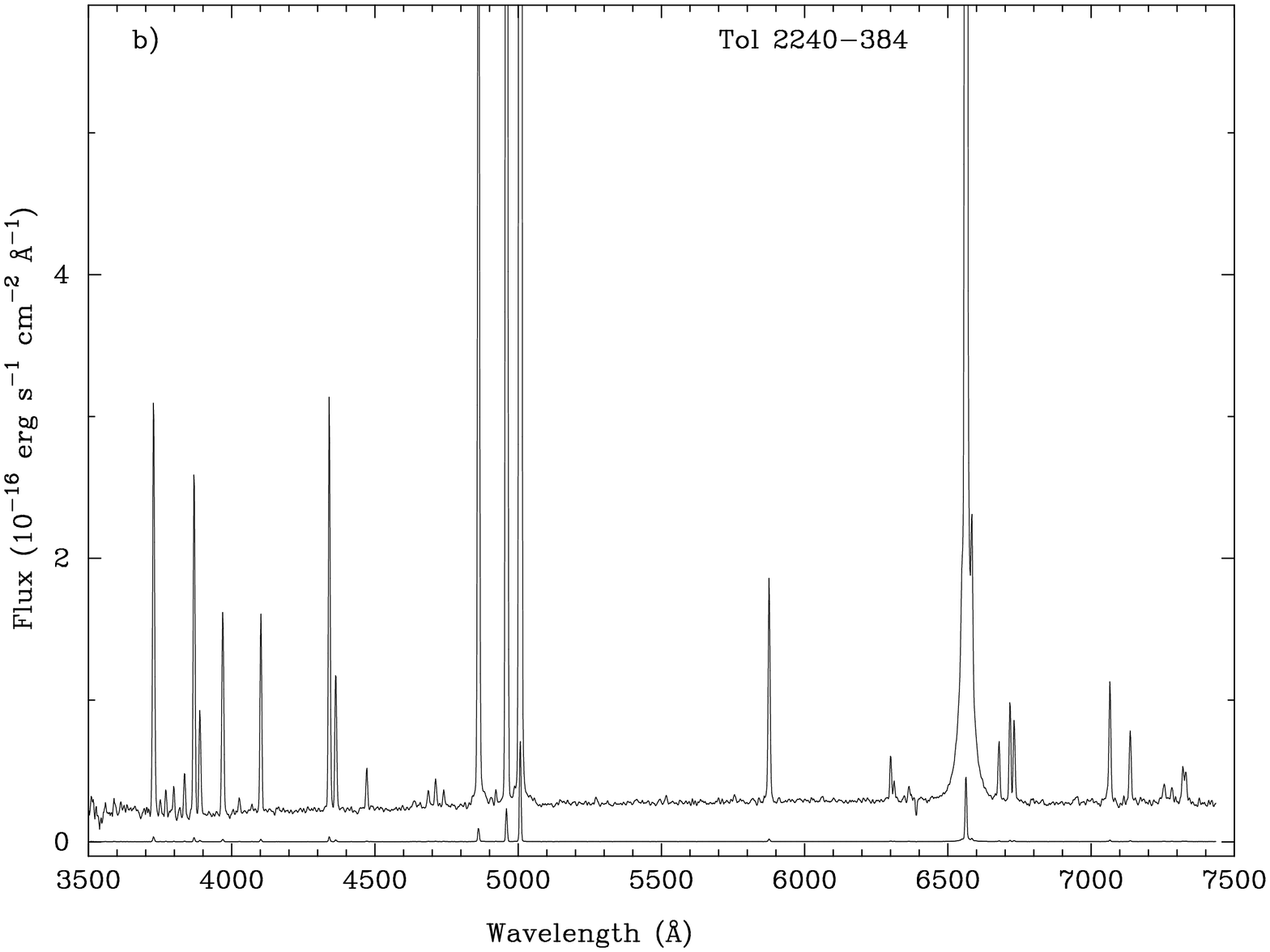,angle=-90,width=7.5cm,clip=}
\caption{Flux-calibrated and redshift-corrected archival VLT/FORS1 
medium-resolution (left) and low-resolution (right) spectra 
of Tol 2240--384 obtained on 12 September 2002 [ESO program 69.C-0203(A)]
(upper spectra).
The lower spectra are the upper spectra downscaled by a factor of 100.
The scale of the ordinate is that for the upper spectra.
Note the strong broad emission in the hydrogen line H$\alpha$ $\lambda$6563
and much weaker broad emission in the hydrogen line H$\beta$ $\lambda$4861.
No appreciable broad emission is detected in strong forbidden lines.}
\label{fig2}
\end{figure*}
%%%%%%%%%%%%%%%%%%%%%%%%%%%%%%%%%%%%%%%%%%%%%%%%%

\section{Observations}

\subsection{Spectroscopy}

A new optical spectrum of Tol 2240--384 was obtained 
using the 8.2 m Very Large Telescope (VLT) on 2009 August 23 
[ESO program 383.B-0271(A)].
The observations were performed using the UVES echelle spectrograph
mounted at the UT2. 
We used the gratings CD\#1 with the central wavelength 3460\AA, 
CD\#2 with the central wavelength 4370\AA, CD\#3 with the central wavelength 
5800\AA, and CD\#4 with the central wavelength 8600\AA.
The slits were used with lengths of 8\arcsec\ and 12\arcsec\
for the blue (CD\#1 and CD\#2) and red (CD\#3 and CD\#4) parts of the spectra,
respectively, and with a width of 3\arcsec.
The angular scale along the slit was 0\farcs246 and 0\farcs182 for the blue
and red arms, respectively.
The above instrumental set-up resulted in
a spectral range $\sim$3000 -- 10 200\AA\ over 131 orders and a 
resolving power $\lambda$/$\Delta$$\lambda$ of
$\sim$ 80 000. The total exposure time was 2960s for gratings CD\#1 and CD\#3,
and 2970s for gratings CD\#2 and CD\#4, divided into 2 equal subexposures.
Observations were performed at airmass $\sim$1.2 with gratings CD\#1 and CD\#3
and $\sim$1.5 with gratings CD\#2 and CD\#4. The seeing was $\sim$2\arcsec.
The Kitt Peak IRS spectroscopic standard star Feige 110 was observed for flux
calibration. Spectra of thorium (Th) comparison arcs were obtained for 
wavelength calibration.

We supplemented the UVES observations with ESO archival data of
Tol 2240--384 (ESO program 69.C-0203(A)). 
These observations were 
obtained on 12 September, 2002 with the FORS1 spectrograph mounted
at the UT3 of the 8.2m ESO VLT.
The observing conditions were photometric throughout the night.

Two sets of spectra were obtained. Low-resolution spectra were
obtained with a grism 300V ($\lambda$$\lambda$$\sim$3850--7500)
and a blocking filter GG 375.
The grisms 600B ($\lambda$$\lambda$$\sim$3560--5970) 
and 600R ($\lambda$$\lambda$$\sim$5330--7480) for the blue 
and red wavelength ranges were used in the medium-resolution observations. To 
avoid second-order contamination, the red part of the spectrum was obtained
with the blocking filter GG 435.

  A long ($\sim$418\arcsec) slit with a width of 0\farcs51 was used.
The spatial scale along the slit was 0\farcs2 pixel$^{-1}$
and the resolving power $\lambda$/$\Delta$$\lambda$ = 300
in the low-resolution mode and
$\lambda$/$\Delta$$\lambda$ = 780 and 1160
in a medium-resolution mode for the 600B and 600R grisms, respectively.
The spectra were obtained at airmass $\sim$ 1.2 -- 1.4.
The seeing was $\sim$ 1\farcs2 during the low-resolution observations,
1\arcsec\ during the medium-resolution observations in the blue range, and 
1\farcs5 during the medium-resolution observations in the red range.
The total integration time for the low-resolution observations was 360s
(3 $\times$ 120s).
The longer exposures were taken for the medium-resolution observations 
and consisted of 
2160s (3 $\times$ 720s) and 1800s (3 $\times$ 600s) 
for the blue and red parts, respectively. 

The two-dimensional spectra were bias subtracted and 
flat-field corrected using IRAF\footnote{IRAF is 
the Image Reduction and Analysis Facility distributed by the 
National Optical Astronomy Observatory, which is operated by the 
Association of Universities for Research in Astronomy (AURA) under 
cooperative agreement with the National Science Foundation (NSF).}. 
We then used the IRAF
software routines IDENTIFY, REIDENTIFY, FITCOORD, and TRANSFORM to 
perform wavelength
calibration and correct for distortion and tilt for each frame. 
 Night sky subtraction was performed using the routine BACKGROUND. 
The level of
night sky emission was determined from the closest regions to the galaxy 
that are free of galaxian stellar and nebular line emission,
 as well as of emission from foreground and background sources.
The one-dimensional spectra were then extracted from the two-dimensional 
frame using the APALL routine. We adopted extraction apertures of  
3\arcsec $\times$ 4\arcsec\ for the UVES spectrum and 
0\farcs51 $\times$ 4\arcsec\ for FORS1 spectra. 
Before extraction, the two 
distinct two-dimensional UVES spectra, the three distinct two-dimensional 
low-resolution FORS1 spectra, and the three distinct two-dimensional 
medium-resolution FORS1 spectra
were carefully aligned with the routine ROTATE using the spatial locations of 
the brightest parts in
each spectrum, so that the spectra were extracted at the same positions in all
subexposures. We then summed the individual spectra 
from each subexposure after manual removal of the cosmic ray hits. 

The resulting UVES spectrum of Tol 2240--384 is shown in Fig. \ref{fig1}. 
A strong broad H$\alpha$ emission line is present in the spectrum, very
similar to the one seen in the spectra of the four low-metallicity AGN
candidates discussed by \citet{I07} and \citet{IT08}. 
The broad component in the H$\beta$ emission line is much weaker,
suggesting a steep Balmer decrement and hence that the 
broad emission originates in a very dense gas.

The extracted medium-resolution and low-resolution spectra of Tol 2240--384
are shown in Figs. \ref{fig2}a and \ref{fig2}b, respectively. As in
the UVES spectrum, a broad H$\alpha$ emission line is detected. 
The signal-to-noise ratio of the FORS1 spectra is higher than that of the 
UVES spectrum because of the lower spectral resolution of the former. 
This allows us to detect broad components of the 
H$\gamma$ and H$\beta$ emission lines.
As in the UVES data, the broad H$\beta$ emission line in the
FORS1 spectra is much weaker than the broad H$\alpha$ emission line.

%%%%%%%%%%%%%%%%%%%%%%%%%%%%%%%%%%%%%%%%%%%%%%%%%%
%   Fig.3 images
%%%%%%%%%%%%%%%%%%%%%%%%%%%%%%%%%%%%%%%%%%%%%%%%%%
\begin{figure*}[t]
\hspace*{0.0cm}\psfig{figure=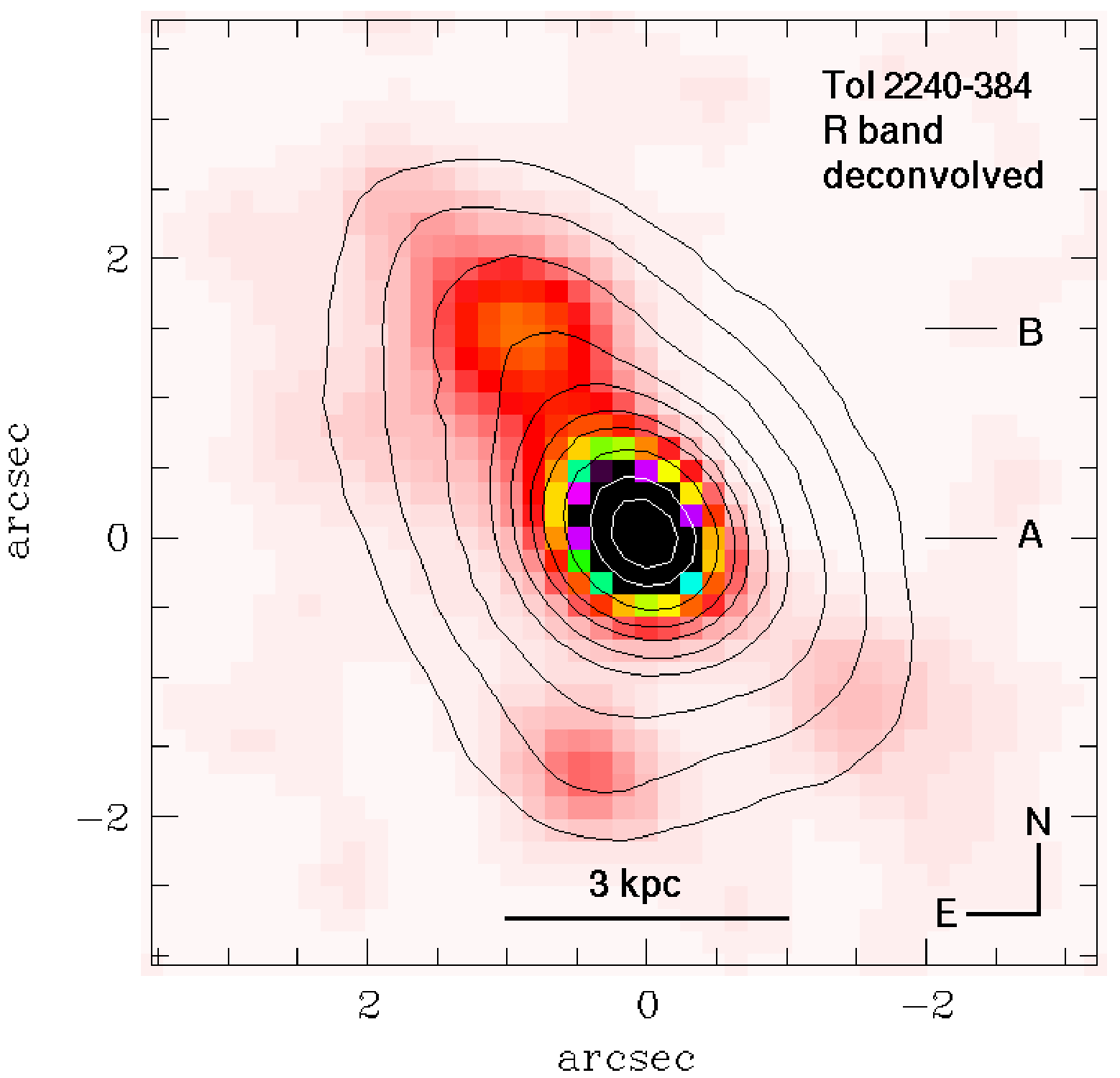,angle=0,height=7cm}%,clip=}
\hspace*{1.0cm}\psfig{figure=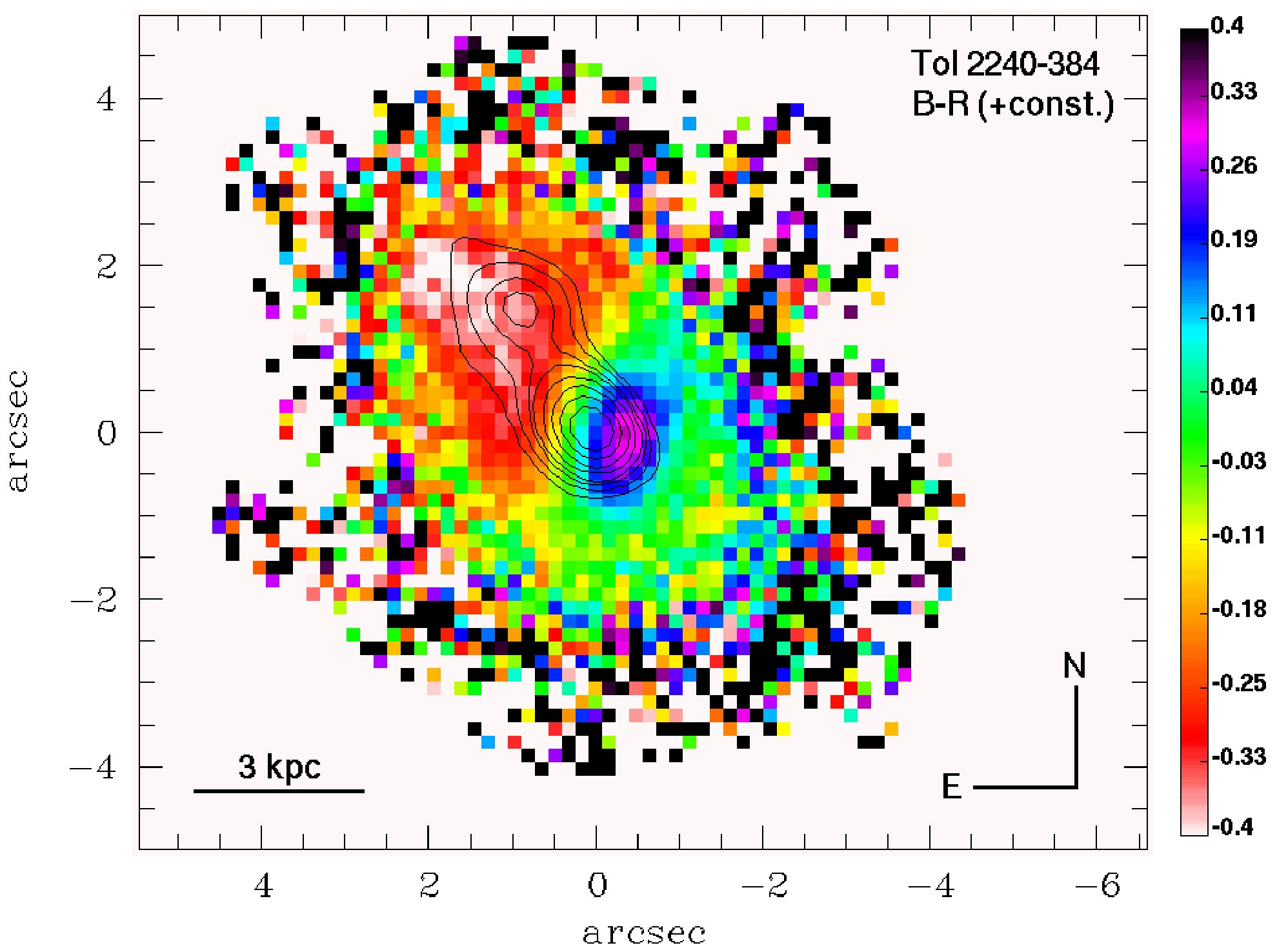,angle=0,height=7cm} %,clip=}
\caption{{\bf left:} Deconvolved $R$ image of Tol~2240--384 revealing two
high-surface brightness regions separated by 1\farcs6 and differing in their
luminosities by a factor $\sim$10. The overlaid contours illustrate the
morphology of the galaxy prior to deconvolution, at a resolution of 0\farcs9 (FWHM).
{\bf right:} Uncalibrated $B-R$ map of Tol~2240--384. Note the presence of
a strong colour contrast between the southwestern and northeastern half  
of the galaxy.
The contours correspond to the deconvolved $R$ image.
}
\label{fig3}
\end{figure*}
%%%%%%%%%%%%%%%%%%%%%%%%%%%%%%%%%%%%%%%%%%%%%%%%%

\subsection{Photometry}

To gain additional insight into the morphological and photometric
properties of Tol 2240--384, we study archival [ESO program 71.B-0509(A)]
images for this system in the Bessel filters $U$ (3 $\times$ 200s),
$B$ (3 $\times$ 400s), and $R$ (3 $\times$ 300s). These data were taken
with the SUSI2 camera (0\farcs161 pixel$^{-1}$) attached to the 3.5m ESO NTT 
in seeing conditions of 
$\approx$1\arcsec\ in $U$ and $B$, and 0\farcs9 in $R$. 
Image reduction and analysis was carried out using MIDAS and additional 
routines developed by ourselves.
From the available calibration exposures, it was not possible to establish
photometric zero points with an accuracy better than $\sim$ 0.2 mag.
In the following, we therefore restrict ourselves to discussing the morphology
and the {\it relative} colour distribution of Tol 2240--384. 
From the approximate apparent $B$ band magnitude of 16 mag that we obtained 
for this system, and assuming a distance of 310 Mpc from the NASA/IPAC Extragalactic Database (NED) 
(corrected for Virgocentric infall and based on $H_0$=73 km s$^{-1}$
Mpc$^{-1}$), we estimate its absolute magnitude to be $\sim$ --21 $B$ mag.
This is 2 -- 4 mag brighter than the absolute SDSS $g$ magnitudes
of four low-metallicity AGNs studied by \citet{IT08}.

\section{Morphology of Tol 2240-384}

The morphology of Tol 2240--384, as inferred from the combined $R$ band
exposure, is illustrated with the contours in 
Fig. \ref{fig3} (left). It can be seen that the galaxy is unresolved and 
shows merely a slight NE--SW elongation on a projected scale of 7$\times$5
kpc. On the same panel, we display the $R$ band image after Lucy deconvolution
\citep{L74}, which contains two main high-surface brightness regions, 
separated by 1\farcs6 (2.4 kpc). 
The southwestern region (labelled A) coincides with the surface 
brightness maximum of the galaxy and is about ten times more luminous than the
northeastern region B. This result was checked and confirmed using
a flux-conserving unsharp masking technique \citep{P98}.

In Fig. \ref{fig3} (right), we show the uncalibrated $B-R$ map
of Tol 2240--384 with the overlaid contours depicting the morphology of the
deconvolved $R$ image. The colour map reveals a strong colour contrast 
of nearly 0.8 mag between the SW and NE half of the galaxy with a relatively 
sharp transition between these two regions at the periphery of knot A.
Knots A and B are located respectively within the red 
($\ga-0.1$ mag) (shown by purple, blue, and green colours in 
Fig. \ref{fig3} (right)) and blue ($\la-0.1$ mag) 
(shown by white and red colours in Fig. \ref{fig3} (right))
halves of Tol 2240--384. They do not, however, spatially coincide with the
locations where the extremal colour indices are observed. 
More specifically, the reddest and bluest features on the colour map are 
offset by between 0\farcs5 and 0\farcs9 from regions A and B. 
The extended, almost uniformly red colour pattern in the SW half of 
Tol 2240--384
is indicative of intense ionised gas emission surrounding the brightest region
A on spatial scales of $\sim$5 kpc. 
Strong H$\alpha$ emission, with an equivalent width of 1300 \AA\ in the UVES
spectrum, 
registered within the $R$ band transmission curve, can readily shift optical
colours by more than 0.5 mag. We note that extreme contamination of optical
colours by extended and intense nebular line emission several kpc away 
from young stellar clusters has been observed in several low-mass starburst 
galaxies \citep[e.g., ][]{I97,P98,P02}. As we discuss below, intense 
H$\alpha$ emission, including a strong broad component, is associated with
region A. In contrast, no appreciable ionised gas emission is present in
the NE part of the galaxy, indicating that the blue $B-R$
colour in region B and its surroundings is mainly due to stellar emission.

\begin{figure}[t]
\hspace*{1.0cm}\psfig{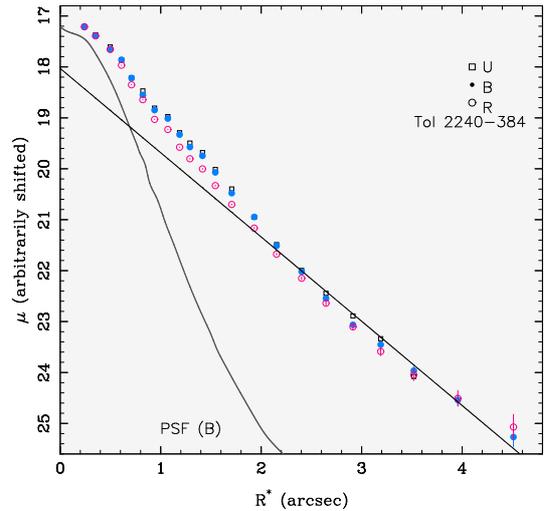}%,clip=}
\caption{ Surface brightness profiles (SBPs) of Tol 2240--384 in $U$,
$B$, and $R$ computed with method iv in \citet{P02}.
The point spread function (PSF) in the $B$ band, derived from two well-exposed
nearby stars in the field of view, is shown with the gray curve.
The straight line shows a linear fit to the host galaxy of Tol 2240--384 
for radii $R^{\star}\geq$1\farcs8.}
\label{fig4}
\end{figure}

In Fig. \ref{fig4}, we show the surface brightness profiles (SBPs) of Tol
2240--384 in $U$ (squares), $B$ (dots) and $R$ (open circles). The SBPs were
computed with the method iv in \citet{P02} and shifted vertically to an equal 
central surface brightness. 
The point spread function (PSF) in the $B$, derived from two well-exposed
nearby stars in the field is included for comparison.
We note that the slight bump in the SBPs at a photometric radius
$R^{\star}\approx$1\farcs6 reflects the luminosity contribution of the
fainter knot B. In agreement with the evidence from the deconvolved $R$ band
image (Fig. \ref{fig3}), the surface photometry does not support the
existence of a bulge component in Tol 2240--384. 
It can be seen that all SBPs have an exponential slope in their outer parts.
This component, reflecting the emission from the host galaxy of 
Tol~2240--384, contains approximately 50\% of the total luminosity
of the galaxy.
The effective radius of 1.2 kpc determined for Tol 2240--384 is a factor 
of between 2 and 3 larger than typical values for blue compact dwarf (BCD) 
galaxies
\citep[cf. ][]{P06}. This is also the case for the exponential scale length 
$\alpha\approx1$ kpc, derived from a linear fit to 
the $B$ band SBP for $R^{\star}\geq$1\farcs8.

%%%%%%%%%%%%%%%%%%%%%%%%%%%%%%%%%%%%%%%%%%%%%%%%%%%%%%%%%%%%%%%%%%%%%%%%%%%%
%   Fig.4 surface brightness profile
%%%%%%%%%%%%%%%%%%%%%%%%%%%%%%%%%%%%%%%%%%%%%%%%%%%%%%%%%%%%%%%%%%%%%%%%%%%
%%%%%%%%%%%%%%%%%%%%%%%%%%%%%%%%%%%%%%%%%%%%%%%%%%%%%%%%%%%%%%%%%%%%%%%%%%

%\renewcommand{\baselinestretch}{1.0}

%%%%%%%%%%%%%%%%%%%%%%%%%%%%%%%%%%%%%%%%%%%%%%%%%%%%%%%%%%%%%%%%%%%%%%%%%%%%
%   Table 1
%%%%%%%%%%%%%%%%%%%%%%%%%%%%%%%%%%%%%%%%%%%%%%%%%%%%%%%%%%%%%%%%%%%%%%%%%%%%
\begin{table*}
\caption{Extinction-corrected narrow emission-line fluxes \label{tab1}}
\centering{
\begin{tabular}{lrrrcrr} \hline\hline
    & \multicolumn{3}{c}{UVES}&&\multicolumn{2}{c}{FORS} \\ \cline{2-4} \cline{6-7}
Line                            &total~~~~~~&blue~~~~~~~&red~~~~~~~&& medium~~         &  low~~~~~~~          \\ \hline %\hline------------------

3188 He {\sc i}                 &3.11$\pm$0.41  &  ...~~~~~~\,  &  ...~~~~~~\,  &&   ...~~~~~~\, &  ...~~~~~~\,  \\
3727 [O {\sc ii}]               &86.95$\pm$1.45 &93.45$\pm$2.13 &69.49$\pm$4.65 &&43.91$\pm$0.74 &64.03$\pm$1.31 \\
3750 H12                        &3.07$\pm$0.49  &  ...~~~~~~\,  &  ...~~~~~~\,  &&3.11$\pm$0.21  &4.59$\pm$0.83  \\
3771 H11                        &3.34$\pm$0.60  &  ...~~~~~~\,  &  ...~~~~~~\,  &&3.61$\pm$0.19  &5.57$\pm$0.62  \\
3797 H10                        &5.26$\pm$0.55  &  ...~~~~~~\,  &  ...~~~~~~\,  &&4.24$\pm$0.18  &5.99$\pm$0.59  \\
3820 He {\sc i}                 &  ...~~~~~~\,  &  ...~~~~~~\,  &  ...~~~~~~\,  &&0.71$\pm$0.07  &1.63$\pm$0.29  \\
3835 H9                         &7.10$\pm$0.05  &  ...~~~~~~\,  &  ...~~~~~~\,  &&5.48$\pm$0.19  &7.91$\pm$0.57  \\
3869 [Ne {\sc iii}]             &53.05$\pm$0.86 &51.50$\pm$0.89 &55.02$\pm$1.36 &&32.69$\pm$0.55 &43.18$\pm$0.87 \\
3889 He {\sc i} + H8            &16.70$\pm$0.72 &  ...~~~~~~\,  &  ...~~~~~~\,  &&11.64$\pm$0.25 &15.53$\pm$0.61 \\
3968 [Ne {\sc iii}] + H7        &34.24$\pm$0.76 &  ...~~~~~~\,  &  ...~~~~~~\,  &&21.62$\pm$0.38 &29.49$\pm$0.71 \\
4026 He {\sc i}                 &2.21$\pm$0.23  &  ...~~~~~~\,  &  ...~~~~~~\,  &&1.16$\pm$0.07  &1.85$\pm$0.24  \\
4069 [S  {\sc ii}]              &  ...~~~~~~\,  &  ...~~~~~~\,  &  ...~~~~~~\,  &&0.79$\pm$0.06  &0.76$\pm$0.16  \\
4076 [S  {\sc ii}]              &  ...~~~~~~\,  &  ...~~~~~~\,  &  ...~~~~~~\,  &&0.28$\pm$0.05  &  ...~~~~~~\,  \\
4101 H$\delta$                  &27.37$\pm$0.66 &27.35$\pm$0.88 &28.91$\pm$1.60 &&18.26$\pm$0.32 &23.34$\pm$0.61 \\
4227 [Fe {\sc v}]               &  ...~~~~~~\,  &  ...~~~~~~\,  &  ...~~~~~~\,  &&2.02$\pm$0.56  &  ...~~~~~~\,  \\
4340 H$\gamma$                  &47.18$\pm$0.87 &47.94$\pm$0.90 &50.30$\pm$1.40 &&37.70$\pm$0.59 &42.98$\pm$0.81 \\
4363 [O {\sc iii}]              &14.73$\pm$0.31 &14.63$\pm$1.09 &14.51$\pm$1.67 &&11.92$\pm$0.20 &14.87$\pm$0.32 \\
4388 He {\sc i}                 &  ...~~~~~~\,  &  ...~~~~~~\,  &  ...~~~~~~\,  &&0.36$\pm$0.04  &  ...~~~~~~\,  \\
4471 He {\sc i}                 &4.46$\pm$0.16  &4.16$\pm$0.42  &4.20$\pm$0.95  &&3.70$\pm$0.08  &4.03$\pm$0.17  \\
4658 [Fe {\sc iii}]             &0.85$\pm$0.09  &  ...~~~~~~\,  &  ...~~~~~~\,  &&0.63$\pm$0.05  &  ...~~~~~~\,  \\
4686 He {\sc ii}                &0.93$\pm$0.09  &  ...~~~~~~\,  &  ...~~~~~~\,  &&1.38$\pm$0.06  &1.76$\pm$0.15  \\
4711 [Ar {\sc iv}] + He {\sc i} &2.52$\pm$0.29  &  ...~~~~~~\,  &  ...~~~~~~\,  &&2.07$\pm$0.06  &2.49$\pm$0.15  \\
4740 [Ar {\sc iv}]              &1.70$\pm$0.09  &  ...~~~~~~\,  &  ...~~~~~~\,  &&1.40$\pm$0.05  &1.52$\pm$0.15  \\
4861 H$\beta$                   &100.00$\pm$1.51&100.00$\pm$1.53&100.00$\pm$1.85&&100.00$\pm$1.49&100.00$\pm$1.60\\
4921 He {\sc i}                 &1.36$\pm$0.11  &  ...~~~~~~\,  &  ...~~~~~~\,  &&0.90$\pm$0.04  &1.06$\pm$0.10  \\
4959 [O {\sc iii}]              &222.99$\pm$3.22&217.93$\pm$3.16&231.89$\pm$3.58&&218.33$\pm$3.24&213.39$\pm$3.29\\
4988 [Fe {\sc iii}]             &0.85$\pm$0.07  &  ...~~~~~~\,  &  ...~~~~~~\,  &&0.75$\pm$0.06  &  ...~~~~~~\,  \\
5007 [O {\sc iii}]              &663.85$\pm$9.54&647.74$\pm$9.35&704.02$\pm$9.99&&663.42$\pm$9.81&644.11$\pm$9.83\\
5016 He {\sc i}                 &2.42$\pm$0.11  &  ...~~~~~~\,  &  ...~~~~~~\,  &&   ...~~~~~~\, &  ...~~~~~~\,  \\
5200 [N {\sc i}]                &0.60$\pm$0.09  &  ...~~~~~~\,  &  ...~~~~~~\,  &&   ...~~~~~~\, &  ...~~~~~~\,  \\
5518 [Cl {\sc iii}]             &0.53$\pm$0.05  &  ...~~~~~~\,  &  ...~~~~~~\,  &&0.32$\pm$0.03  &  ...~~~~~~\,  \\
5755 [N {\sc ii}]               &0.48$\pm$0.13  &  ...~~~~~~\,  &  ...~~~~~~\,  &&0.38$\pm$0.05  &0.48$\pm$0.08  \\
5876 He {\sc i}                 &13.76$\pm$0.23 &13.83$\pm$0.29 &13.96$\pm$0.76 &&12.51$\pm$0.22 &12.99$\pm$0.25 \\
6300 [O {\sc i}]                &2.58$\pm$0.06  &  ...~~~~~~\,  &  ...~~~~~~\,  &&2.01$\pm$0.06  &2.13$\pm$0.08  \\
6312 [S {\sc iii}]              &1.19$\pm$0.04  &  ...~~~~~~\,  &  ...~~~~~~\,  &&0.88$\pm$0.04  &0.87$\pm$0.07  \\
6364 [O {\sc i}]                &0.78$\pm$0.04  &  ...~~~~~~\,  &  ...~~~~~~\,  &&0.62$\pm$0.04  &0.72$\pm$0.07  \\
6548 [N {\sc ii}]               &2.09$\pm$0.06  &  ...~~~~~~\,  &  ...~~~~~~\,  &&   ...~~~~~~\, &  ...~~~~~~\,  \\
6563 H$\alpha$                  &279.49$\pm$4.36&279.38$\pm$4.38&279.85$\pm$4.65&&284.13$\pm$4.56&282.54$\pm$4.69\\
6583 [N {\sc ii}]               &6.08$\pm$0.10  &  ...~~~~~~\,  &  ...~~~~~~\,  &&5.68$\pm$0.12  &5.01$\pm$0.14  \\
6678 He  {\sc i}                &3.12$\pm$0.07  &  ...~~~~~~\,  &  ...~~~~~~\,  &&2.77$\pm$0.07  &2.51$\pm$0.09  \\
6716 [S {\sc ii}]               &5.67$\pm$0.12  &  ...~~~~~~\,  &  ...~~~~~~\,  &&4.10$\pm$0.09  &3.75$\pm$0.10  \\
6731 [S {\sc ii}]               &4.41$\pm$0.08  &  ...~~~~~~\,  &  ...~~~~~~\,  &&3.56$\pm$0.08  &3.05$\pm$0.09  \\
7065 He  {\sc i}                &1.45$\pm$0.05  &  ...~~~~~~\,  &  ...~~~~~~\,  &&  ...~~~~~~\,  &4.60$\pm$0.12  \\
7136 [Ar {\sc iii}]             &4.13$\pm$0.12  &  ...~~~~~~\,  &  ...~~~~~~\,  &&  ...~~~~~~\,  &2.41$\pm$0.09  \\
7281 He {\sc i}                 &0.81$\pm$0.04  &  ...~~~~~~\,  &  ...~~~~~~\,  &&   ...~~~~~~\, &  ...~~~~~~\,  \\
7320 [O {\sc ii}]               &2.40$\pm$0.09  &  ...~~~~~~\,  &  ...~~~~~~\,  &&  ...~~~~~~\,  &  ...~~~~~~\,  \\
7330 [O {\sc ii}]               &1.78$\pm$0.06  &  ...~~~~~~\,  &  ...~~~~~~\,  &&  ...~~~~~~\,  &  ...~~~~~~\,  \\
7751 [Ar {\sc iii}]             &0.81$\pm$0.04  &  ...~~~~~~\,  &  ...~~~~~~\,  &&   ...~~~~~~\, &  ...~~~~~~\,  \\
8446 O {\sc i}                  &1.02$\pm$0.04  &  ...~~~~~~\,  &  ...~~~~~~\,  &&   ...~~~~~~\, &  ...~~~~~~\,  \\
8467 P17                        &0.39$\pm$0.02  &  ...~~~~~~\,  &  ...~~~~~~\,  &&   ...~~~~~~\, &  ...~~~~~~\,  \\
8502 P16                        &0.50$\pm$0.03  &  ...~~~~~~\,  &  ...~~~~~~\,  &&   ...~~~~~~\, &  ...~~~~~~\,  \\
8545 P15                        &0.44$\pm$0.03  &  ...~~~~~~\,  &  ...~~~~~~\,  &&   ...~~~~~~\, &  ...~~~~~~\,  \\
8598 P14                        &0.70$\pm$0.03  &  ...~~~~~~\,  &  ...~~~~~~\,  &&   ...~~~~~~\, &  ...~~~~~~\,  \\
8750 P12                        &1.21$\pm$0.04  &  ...~~~~~~\,  &  ...~~~~~~\,  &&   ...~~~~~~\, &  ...~~~~~~\,  \\
8863 P11                        &1.60$\pm$0.05  &  ...~~~~~~\,  &  ...~~~~~~\,  &&   ...~~~~~~\, &  ...~~~~~~\,  \\
9015 P10                        &2.03$\pm$0.05  &  ...~~~~~~\,  &  ...~~~~~~\,  &&   ...~~~~~~\, &  ...~~~~~~\,  \\
9069 [S {\sc iii}]              &8.10$\pm$0.20  &  ...~~~~~~\,  &  ...~~~~~~\,  &&   ...~~~~~~\, &  ...~~~~~~\,  \\ %\\
$C$(H$\beta$)                   &\multicolumn{1}{c}{0.280}&\multicolumn{1}{c}{0.290}&\multicolumn{1}{c}{0.195}&&\multicolumn{1}{c}{0.830}&\multicolumn{1}{c}{0.815} \\
EW(H$\beta$)$^a$                &\multicolumn{1}{c}{167}&\multicolumn{1}{c}{112}&\multicolumn{1}{c}{46}&&\multicolumn{1}{c}{284}&\multicolumn{1}{c}{274} \\
$F$(H$\beta$)$^b$               &\multicolumn{1}{c}{234}&\multicolumn{1}{c}{157}&\multicolumn{1}{c}{65}&&\multicolumn{1}{c}{80}&\multicolumn{1}{c}{73} \\
EW(abs)$^a$                     &\multicolumn{1}{c}{0.2}&\multicolumn{1}{c}{0.3}&\multicolumn{1}{c}{0.1}&&\multicolumn{1}{c}{5.2}&\multicolumn{1}{c}{5.6} \\
\hline \\
$^a$ in \AA. \\
$^b$ in units 10$^{-16}$ erg s$^{-1}$ cm$^{-2}$. 
\end{tabular}
}

\end{table*}
%%%%%%%%%%%%%%%%%%%%%%%%%%%%%%%%%%%%%%%%%%%%%%%%%%%%%%%%%%%%%%%%%%%%%%%%%%%%%%%%%%%%%%

%\renewcommand{\baselinestretch}{1.5}

%\renewcommand{\baselinestretch}{1.0}

%%%%%%%%%%%%%%%%%%%%%%%%%%%%%%%%%%%%%%%%%%%%%%%%%%%%%%%%%%%%%%%%%%%%%%%%%%%%%%%%%%%%%%
%    Table 2
%%%%%%%%%%%%%%%%%%%%%%%%%%%%%%%%%%%%%%%%%%%%%%%%%%%%%%%%%%%%%%%%%%%%%%%%%%%%%%%%%%%%%%
\begin{table*}[t]
\caption{Physical conditions and element abundances \label{tab2}}
%\centering{
\begin{tabular}{lcccccc} \hline\hline
    & \multicolumn{3}{c}{UVES}&&\multicolumn{2}{c}{FORS} \\ \cline{2-4} \cline{6-7}
Property                            &total&blue&red&& medium         &  low          \\ \hline %\hline

$T_e$(O {\sc iii}), K                  &15980$\pm$190  &16110$\pm$590  &15470$\pm$840  &&14540$\pm$130  &15780$\pm$200   \\
$T_e$(O {\sc ii}), K                   &14860$\pm$160  &14930$\pm$500  &14560$\pm$720  &&13960$\pm$120  &14740$\pm$170   \\
$T_e$(S {\sc iii}), K                  &14960$\pm$150  &15070$\pm$490  &15010$\pm$700  &&14250$\pm$110  &14800$\pm$170   \\
$N_e$(O {\sc ii}), cm$^{-3}$           & 300$\pm$50    &  350$\pm$50   &  150$\pm$140  &&  100$\pm$50   &     ...        \\
$N_e$(S {\sc ii}), cm$^{-3}$           & 140$\pm$40    &   ...         &   ...         &&  320$\pm$60   &  210$\pm$70    \\ \\
O$^+$/H$^+$, ($\times$10$^5$)          &0.82$\pm$0.03  &0.86$\pm$0.08  &0.69$\pm$0.10  &&0.52$\pm$0.02  &0.63$\pm$0.02   \\
O$^{2+}$/H$^+$, ($\times$10$^5$)       &6.26$\pm$0.20  &5.99$\pm$0.55  &7.15$\pm$0.99  &&7.87$\pm$0.21  &6.24$\pm$0.21   \\
O$^{3+}$/H$^+$, ($\times$10$^6$)       &0.59$\pm$0.06  &       ...     &       ...     &&1.27$\pm$0.07  &1.20$\pm$0.11   \\
O/H, ($\times$10$^5$)                  &7.14$\pm$0.20  &6.85$\pm$0.56  &7.78$\pm$1.00  &&8.52$\pm$0.21  &6.99$\pm$0.21   \\
12+log O/H                             &7.85$\pm$0.01  &7.84$\pm$0.04  &7.89$\pm$0.06  &&7.93$\pm$0.01  &7.84$\pm$0.01   \\ \\
N$^{+}$/H$^+$, ($\times$10$^6$)        &0.46$\pm$0.01  &      ...      &      ...      &&0.49$\pm$0.01  &0.38$\pm$0.01   \\
$ICF$(N)$^a$                               &      8.20     &      ...      &      ...      &&     15.1      &     10.4       \\
N/H, ($\times$10$^6$)                  &3.74$\pm$0.09  &      ...      &      ...      &&7.33$\pm$0.18  &3.99$\pm$0.13   \\
log N/O                                &--1.28$\pm$0.02~~&      ...      &      ...      &&--1.07$\pm$0.02~~&--1.24$\pm$0.02~~ \\ \\
Ne$^{2+}$/H$^+$, ($\times$10$^5$)      &1.19$\pm$0.04  &1.13$\pm$0.11  &1.35$\pm$0.19  &&0.96$\pm$0.03  &1.00$\pm$0.04   \\
$ICF$(Ne)$^a$                              &      1.04     &      1.04     &      1.03     &&      1.03     &      1.04      \\
Ne/H, ($\times$10$^5$)                 &1.23$\pm$0.05  &1.17$\pm$0.12  &1.39$\pm$0.21  &&0.99$\pm$0.03  &1.04$\pm$0.04   \\
log Ne/O                               &--0.76$\pm$0.02~~&--0.77$\pm$0.06~~&--0.75$\pm$0.09~~&&--0.94$\pm$0.02~~&--0.83$\pm$0.02~~ \\ \\
S$^{+}$/H$^+$, ($\times$10$^6$)        &0.10$\pm$0.01  &       ...     &       ...     &&0.09$\pm$0.01  &0.07$\pm$0.01   \\
S$^{2+}$/H$^+$, ($\times$10$^6$)       &0.62$\pm$0.03  &       ...     &       ...     &&0.53$\pm$0.03  &0.47$\pm$0.04   \\
$ICF$(S)$^a$                               &      1.87     &       ...     &       ...     &&      3.02     &      2.24      \\
S/H, ($\times$10$^6$)                  &1.34$\pm$0.05  &       ...     &       ...     &&1.87$\pm$0.08  &1.20$\pm$0.09   \\
log S/O                                &--1.73$\pm$0.02~~&       ...     &       ...     &&--1.66$\pm$0.02~~&--1.76$\pm$0.03~~ \\ \\
Cl$^{2+}$/H$^+$, ($\times$10$^8$)      &2.43$\pm$0.19  &       ...     &       ...     &&1.65$\pm$0.13  &       ...      \\
$ICF$(Cl)$^a$                              &      1.28     &       ...     &       ...     &&      1.64     &       ...      \\
Cl/H, ($\times$10$^8$)                 &3.12$\pm$0.24  &       ...     &       ...     &&2.71$\pm$0.21  &       ...      \\
log Cl/O                               &--3.36$\pm$0.04~~&       ...     &       ...     &&--3.50$\pm$0.03~~&       ...      \\ \\
Ar$^{2+}$/H$^+$, ($\times$10$^7$)      &1.67$\pm$0.05  &       ...     &       ...     &&      ...      &0.99$\pm$0.04   \\
Ar$^{3+}$/H$^+$, ($\times$10$^7$)      &      ...      &       ...     &       ...     &&1.54$\pm$0.07  &1.35$\pm$0.14   \\
$ICF$(Ar)$^a$                              &      1.80     &       ...     &       ...     &&       ...     &      2.16      \\
Ar/H, ($\times$10$^7$)                 &2.99$\pm$0.10  &       ...     &       ...     &&      ...      &2.14$\pm$0.31   \\
log Ar/O                               &--2.38$\pm$0.02~~&       ...     &       ...     &&      ...      &--2.51$\pm$0.06~~ \\ \\
Fe$^{2+}$/H$^+$, ($\times$10$^6$)(4658)&0.17$\pm$0.02  &       ...     &       ...     &&0.14$\pm$0.01  &       ...      \\
$ICF$(Fe)                              &      12.0     &       ...     &       ...     &&      22.8     &       ...      \\
Fe/H, ($\times$10$^6$)(4658)           &1.97$\pm$0.20  &       ...     &       ...     &&3.27$\pm$0.26  &       ...      \\
log Fe/O (4658)                        &--1.22$\pm$0.05~~&       ...     &       ...     &&--1.00$\pm$0.03~~&       ...      \\ \\
Fe$^{2+}$/H$^+$, ($\times$10$^6$)(4988)&       ...     &       ...     &       ...     &&0.17$\pm$0.01  &       ...      \\
$ICF$(Fe)$^a$                              &       ...     &       ...     &       ...     &&      22.8     &       ...      \\
Fe/H, ($\times$10$^6$)(4988)           &       ...     &       ...     &       ...     &&3.88$\pm$0.32  &       ...      \\
log Fe/O (4988)                        &       ...     &       ...     &       ...     &&--0.92$\pm$0.04~~&       ...      \\
\hline
\end{tabular}

\smallskip

$^a$ Ionisation correction factor.
%}
\end{table*}
%%%%%%%%%%%%%%%%%%%%%%%%%%%%%%%%%%%%%%%%%%%%%%%%%%%%%%%%%%%%%%%%%%%%%%%%%%%%%%%%%%%%%%

%\renewcommand{\baselinestretch}{1.5}

%%%%%%%%%%%%%%%%%%%%%%%%%%%%%%%%%%%%%%%%%%%%%%%%
%    Fig.5  (Diagnostic diagram)
%%%%%%%%%%%%%%%%%%%%%%%%%%%%%%%%%%%%%%%%%%%%%%%%
\begin{figure*}[t]
\hspace*{2.5cm}\psfig{figure=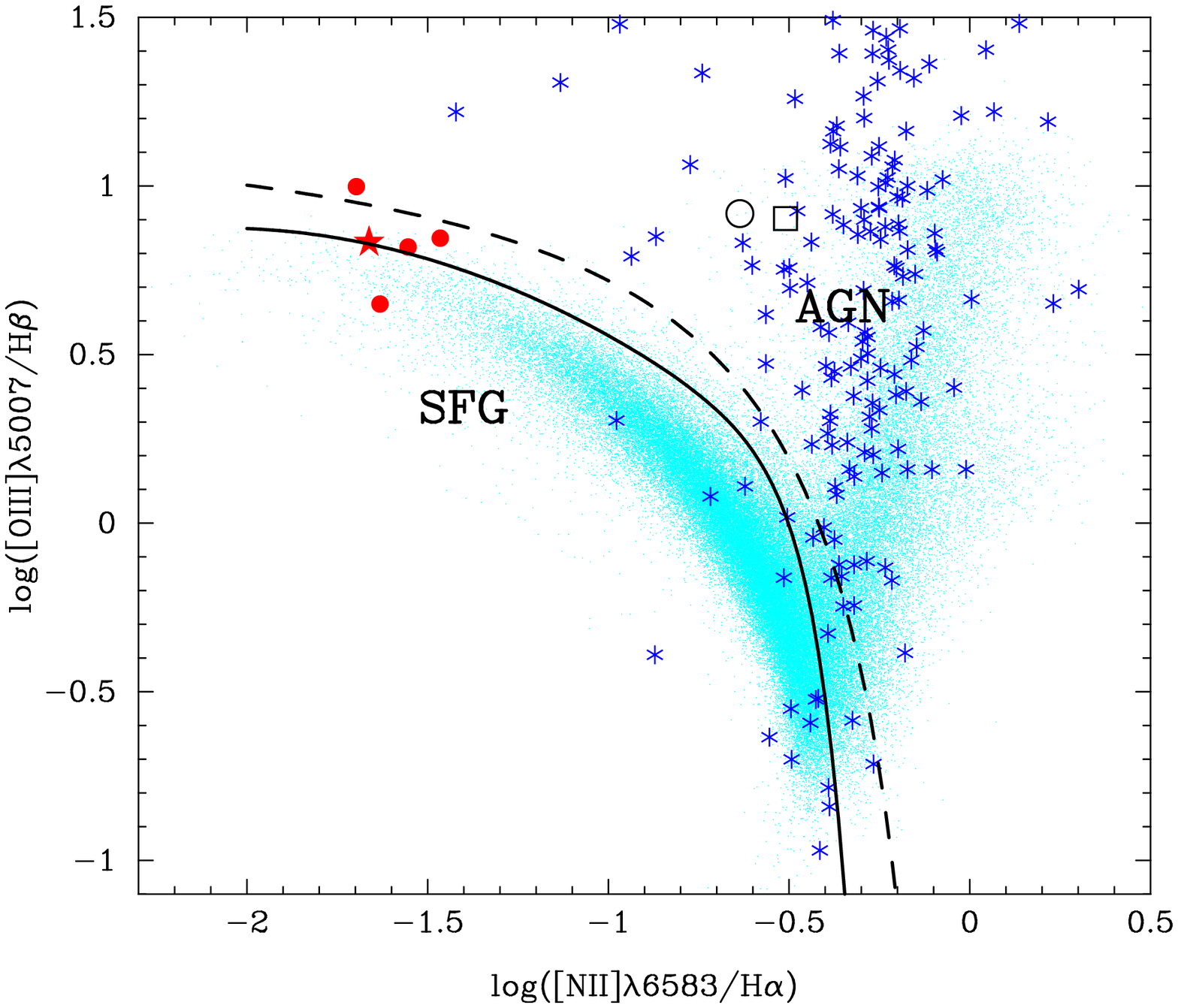,angle=-90,width=12.cm,clip=}
\caption{The Baldwin-Phillips-Terlevich (BPT) diagram \citep{B81} for narrow
emission lines. Plotted are the $\sim$100 000 ELGs
  from SDSS DR7 (cloud of dots), the 
SDSS sample of broad-line AGNs with low-mass black holes of 
\citet{G07} (asterisks), the low-luminosity broad-line AGNs 
NGC 4395 \citep[open circle, ][]{K99} and Pox 52 \citep[open square, ][]{B04},
the four low-metallicity AGNs from \citet{IT08}
(filled circles) and Tol 2240--384 (star). The dashed line by \citet{K03}
and the solid line by \citet{S06} separate 
star-forming galaxies (SFG) from active galactic nuclei (AGN). 
}
\label{fig5}
\end{figure*}
%%%%%%%%%%%%%%%%%%%%%%%%%%%%%%%%%%%%%%%%%%%%%%%%%

%%%%%%%%%%%%%%%%%%%%%%%%%%%%%%%%%%%%%%%%%%%%%%%%
%    Fig.6  (decomposition of narrow lines)
%%%%%%%%%%%%%%%%%%%%%%%%%%%%%%%%%%%%%%%%%%%%%%%%
\begin{figure*}[t]
\hspace*{2.5cm}\psfig{figure=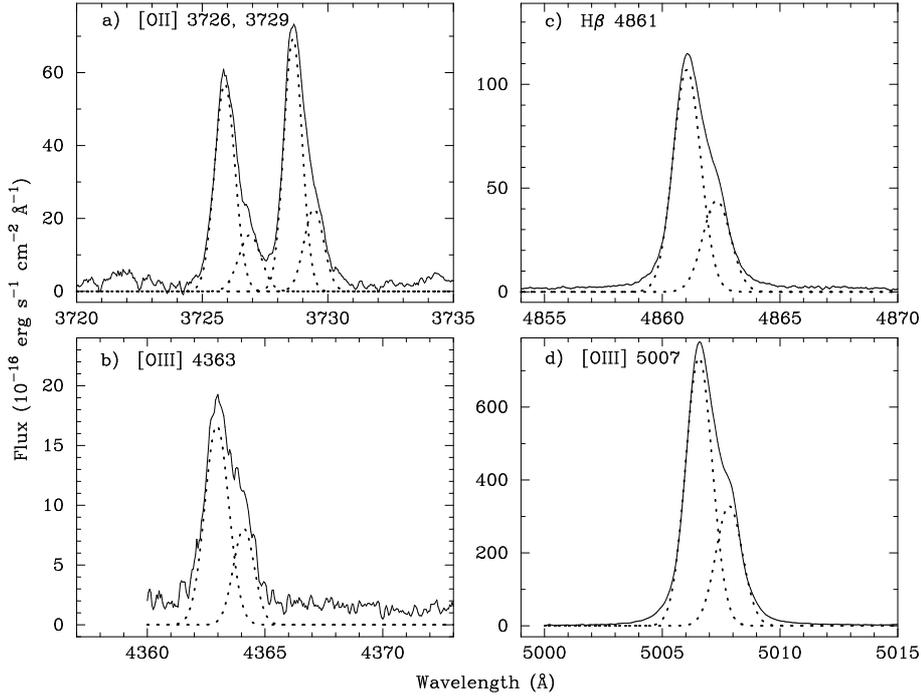,angle=-90,width=12.cm}%,clip=}
\caption{Decomposition of strong emission line profiles into two Gaussian 
components in the UVES spectrum of Tol 2240--384 for:
a) [O~{\sc ii}] $\lambda$3726 and $\lambda$3729, 
b) [O~{\sc iii}] $\lambda$4363,
c) H$\beta$ $\lambda$4861 and d) [O~{\sc iii}] $\lambda$5007. 
Observational data are shown by solid lines and fits by dotted lines.}
\label{fig6}
\end{figure*}
%%%%%%%%%%%%%%%%%%%%%%%%%%%%%%%%%%%%%%%%%%%%%%%%%

%%%%%%%%%%%%%%%%%%%%%%%%%%%%%%%%%%%%%%%%%%%%%%%%
%    Fig.7 (FORS)
%%%%%%%%%%%%%%%%%%%%%%%%%%%%%%%%%%%%%%%%%%%%%%%%
\begin{figure*}[t]
\hspace*{2.5cm}\psfig{figure=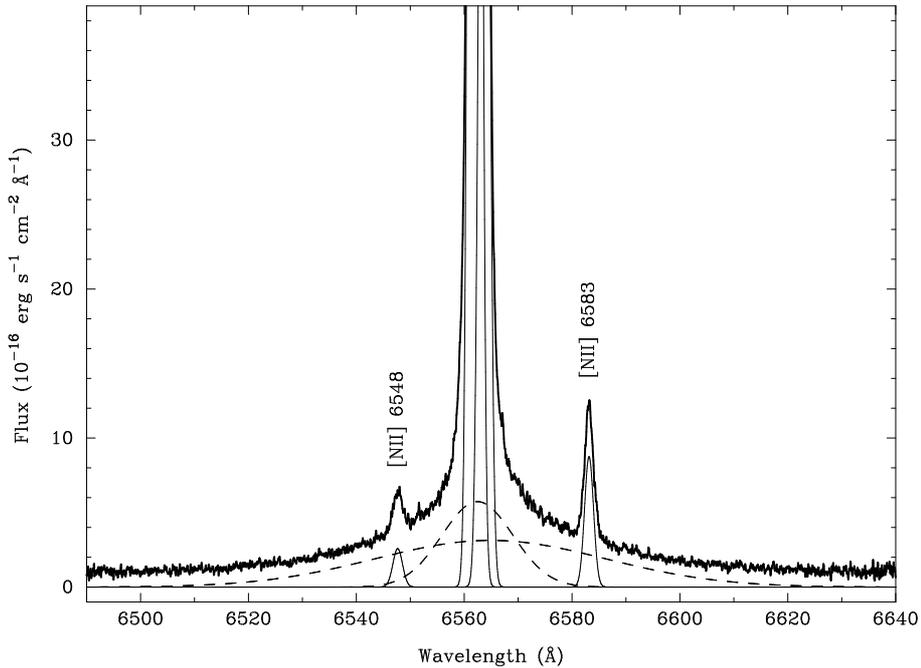,angle=-90,width=12.cm}%,clip=}
\caption{Decomposition of the H$\alpha$ $\lambda$6563 emission line profile 
into four Gaussian components. Two broad components of 
the H$\alpha$ emission line
are shown by dashed lines, while two narrow components of H$\alpha$ and
fits by single Gaussians for the [N {\sc ii}] $\lambda$6548 and $\lambda$6583
emission lines are shown by solid lines.}
\label{fig7}
\end{figure*}
%%%%%%%%%%%%%%%%%%%%%%%%%%%%%%%%%%%%%%%%%%%%%%%%%

%%%%%%%%%%%%%%%%%%%%%%%%%%%%%%%%%%%%%%%%%%%%%%%%
%    Fig.8  (Ha/Hb ratio vs density)
%%%%%%%%%%%%%%%%%%%%%%%%%%%%%%%%%%%%%%%%%%%%%%%%
\begin{figure*}[t]
\hspace*{2.5cm}\psfig{figure=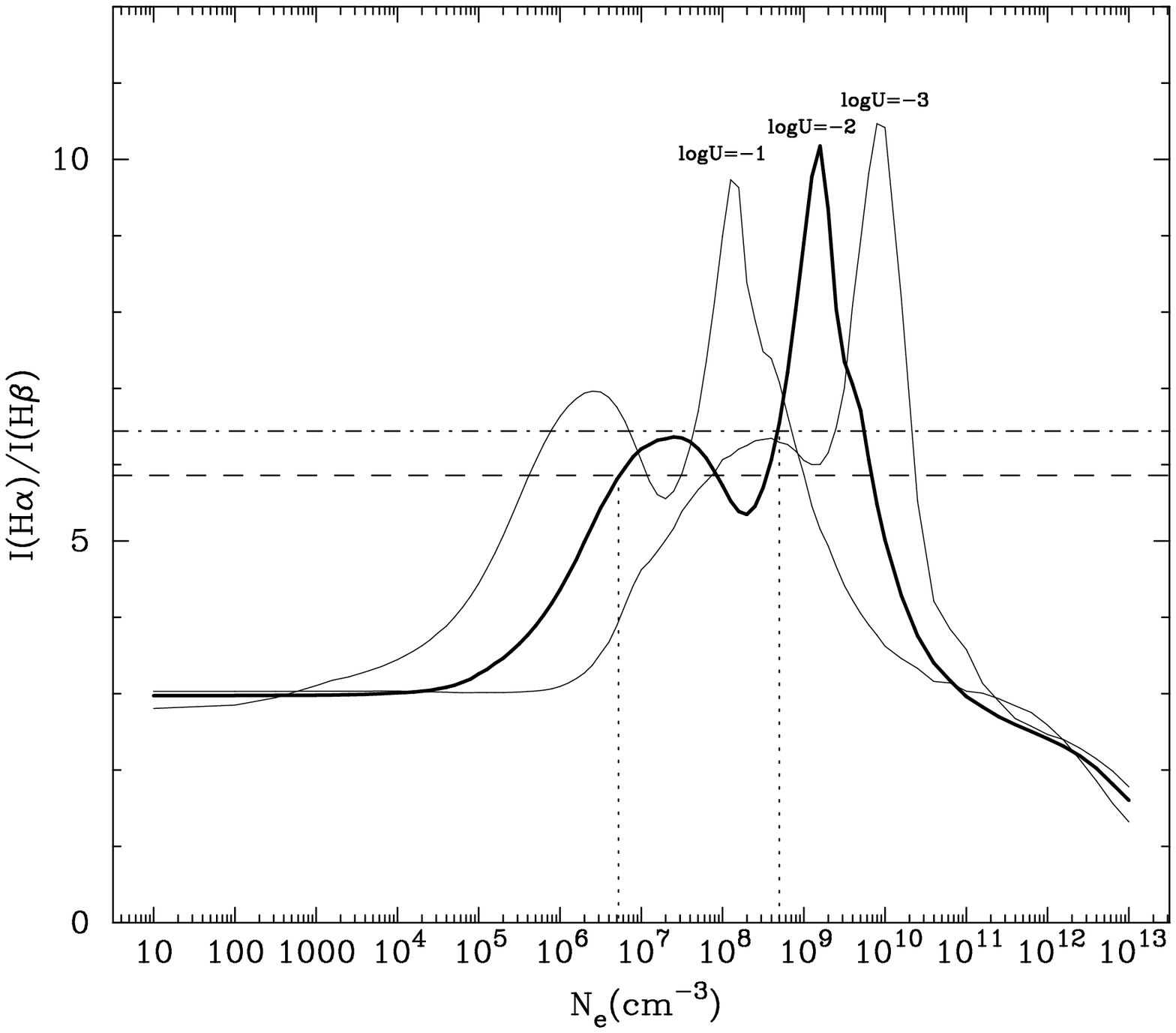,angle=-90,width=12.cm,clip=}
\caption{The dependence of the theoretical H$\alpha$-to-H$\beta$ emission line
flux ratio $I$(H$\alpha$)/$I$(H$\beta$) on the electron number density 
$N_e$ calculated with the CLOUDY photoionisation models for three
values of the ionisation parameter log $U$ = --2 (thick solid line),
log $U$ = --1 and log $U$ = --3 (thin solid lines) \citep{F96,F98}.
For the ionising radiation, the power law spectral energy distribution 
$f_\nu$ $\propto$ $\nu^\alpha$ for $\alpha$ = --1 is chosen.
An oxygen abundance of 12+logO/H = 7.6 is adopted for the gas. 
The dash-dotted line is the 
extinction-corrected H$\alpha$-to-H$\beta$ flux ratio measured in the UVES 
spectrum adopting $C$(H$\beta$) = 0.28, and the dashed line is the 
extinction-corrected H$\alpha$-to-H$\beta$ flux ratio measured in the 
medium-resolution FORS1 spectrum adopting the extinction coefficient
$C$(H$\beta$) = 0.83.
The range of possible electron number densities in the region of broad 
emission is shown by two vertical dotted lines assuming an ionising radiation
field with log $U$ = --2.}
\label{fig8}
\end{figure*}
%%%%%%%%%%%%%%%%%%%%%%%%%%%%%%%%%%%%%%%%%%%%%%%%%

%\section{Results and discussion}

\section{Location in the emission-line diagnostic diagram}

Figure \ref{fig5} shows the position of Tol 2240--384 (represented by a star) 
in the classical  [O~{\sc iii}] $\lambda$5007/H$\beta$ vs.  
[N {\sc ii}] $\lambda$6583/H$\alpha$ diagram \citep[][hereafter BPT]{B81}, 
in addition to other objects shown for comparison. The four objects from 
\citet{IT08} are represented by filled circles and lie in the same region. 
The asterisks represent the broad-line AGNs with low black hole masses from 
\citet{G07}. The  two low-luminosity broad-line AGNs NGC 4395 and Pox 52 
\citep{K99,B04} are represented by an open circle and an open square, 
respectively. For reference, the cyan dots represent all the galaxies 
from the SDSS DR7 with flux errors smaller than 10\% for each of the four 
emission lines, H$\beta$, [O~{\sc iii}] $\lambda$5007, H$\alpha$ and
[N {\sc ii}] $\lambda$6583. The emission line fluxes were measured using the 
technique developed by \citet{T04} and were taken from the SDSS 
website.\footnote{http://www.sdss.org/DR7/products/value$_-$added/index.html.} 
These galaxies are distributed into two wings, the left one interpreted as 
star-forming galaxies and the right one containing AGN hosts. The dashed line 
represents the empirical divisory line between star-forming galaxies and AGNs 
drawn by \citet{K03}, while the continuous line represents the upper limit 
for pure star-forming galaxies from \citet{S06}. One can see that  
Tol 2240--384, as well as the four objects from \cite{IT08} lie in the 
low metallicity part of what is usually considered as the region of 
star-forming galaxies. However, it has been shown by \citet{S06} that an AGN 
hosted by a low metallicity galaxy would be difficult to distinguish in this 
diagram, even if  the active nucleus contributed significantly to the 
emission lines (which is not the case in Tol 2240--384, as discussed later in 
this paper). The active  galaxies from the \citet{G07} sample  occupy a 
different zone in the BPT diagram, closer to the ``AGN wing'', probably
because their metallicities are higher than that of Tol 2240--384 and the 
four galaxies from \citet{IT08}, but lower than those of the bulk of AGN
hosts. 

\section{Element abundances}

\subsection{Empirical analysis}

We derived element abundances from the narrow emission-line fluxes, 
using a classical empirical method.
Thanks to the high spectral resolution, the narrow emission lines in the UVES 
spectrum are resolved and found to have an
asymmetric shape, implying the presence of two kinematically distinct
emission-line regions in the SW part of the galaxy
(Fig. \ref{fig6}). On the other hand,
these lines are not resolved in the FORS1 spectra due to insufficient
spectral resolution and have therefore the full widths at half maximum 
(FWHM) corresponding to the instrumental ones.
Therefore, in the case of the
UVES spectrum, the narrow emission lines were deblended and element abundances
were derived from the emission line fluxes for every emitting region.
To improve the accuracy of the abundance determination, we also derived 
element abundances for the total emission-line fluxes, including both
blueshifted and redshifted components of the emission lines.
The fluxes in all spectra were 
measured using Gaussian fitting with the IRAF SPLOT routine. 
They were corrected for both extinction, using the reddening curve
of \citet{W58}, and underlying
hydrogen stellar absorption, derived simultaneously by an iterative procedure 
described by \citet{ITL94} and using the observed decrements of the 
narrow hydrogen Balmer lines. The extinction coefficient 
$C$(H$\beta$) and equivalent width of hydrogen absorption lines EW(abs) are 
derived in such a way to obtain the closest agreement between the 
extinction-corrected and theoretical recombination hydrogen emission-line 
fluxes normalised to the H$\beta$ flux. It is assumed that EW(abs) is the same
for all hydrogen lines. This assumption is justified by the evolutionary 
stellar population synthesis models of \citet{GD05}.

The extinction-corrected total fluxes 
100$\times$$I$($\lambda$)/$I$(H$\beta$) of the narrow lines from 
the UVES spectrum as well as fluxes of the blueshifted
and redshifted components, and the extinction coefficient
$C$(H$\beta$), the equivalent width of the H$\beta$ emission line 
EW(H$\beta$), the H$\beta$ observed flux $F$(H$\beta$), and the 
equivalent width of the 
underlying hydrogen absorption lines EW(abs) are given in Table \ref{tab1}.
We note that total fluxes of hydrogen emission lines corrected for 
extinction and underlying hydrogen absorption 
(column 2 in Table \ref{tab1}) are very 
close to the theoretical recombination values 
of \citet{HS87} suggesting that the extinction
coefficient $C$(H$\beta$) is reliably derived. We obtained  EW(abs)
of $\sim$ 0.2\AA, 
which is much smaller than the equivalent widths of hydrogen emission lines,
implying that the effect of underlying absorption on the emission line
fluxes is very small, $\sim$ 2 percent for H9 $\lambda$3835 and much lower
for stronger lines.
In Table \ref{tab1}, we also show the emission-line fluxes and other 
parameters for the medium- and 
low-resolution FORS1 spectra. We note that the extinction coefficient 
$C$(H$\beta$) (Table \ref{tab1}) derived from the FORS1 spectra is 
significantly higher than that derived from the UVES spectrum. 
This difference is probably caused by the FORS1 spectra being obtained 
at the relatively high airmass of $\sim$ 1.2 -- 1.4 with the 
narrow 0\farcs51 slit. Therefore,
these spectra are affected by atmospheric refraction. This effect
is seen by comparing the continuum slopes of the UVES and FORS1
spectra (Figs. \ref{fig1} and \ref{fig2}), respectively. 
The continuum in the UVES spectrum
is blue, while it is reddish in the FORS1 spectra. This effect is
somewhat larger for the medium-resolution spectrum.
We conclude that the FORS1 data are somewhat uncertain for the analysis of 
physical conditions and the abundance determination.

%%%%%%%%%%%%%%%%%%%%%%%%%%%%%%%%%%%%%%%%%%%%%%%%%%%%%%%%%%%%%%%%%%%%%%%%%%%%
%   Table 3 (models) 
%%%%%%%%%%%%%%%%%%%%%%%%%%%%%%%%%%%%%%%%%%%%%%%%%%%%%%%%%%%%%%%%%%%%%%%%%%%%
\begin{table} [t]
\caption{Comparison of photoionisation models with observations}
 \label{tab3}
\centering{
\begin{tabular}{lrrr} \hline \hline
Line                            &observed&model M1 & model M2\\ \hline %\hline
3727 [O {\sc ii}]               &86.95$\pm$1.45 & 86.58 & 86.21 \\
3869 [Ne {\sc iii}]             &53.05$\pm$0.86 &53.30 & 53.34\\
%4069 [S  {\sc ii}]              &  ...~~~~~~\,  &\\
%4076 [S  {\sc ii}]              &  ...~~~~~~\,  &\\
4363 [O {\sc iii}]              &14.73$\pm$0.31 & 14.55 & 14.67\\
4471 He {\sc i}                 &4.46$\pm$0.16  &  5.10 & 5.00\\
4658 [Fe {\sc iii}]             &0.85$\pm$0.09  & 0.86 & 0.85\\
4686 He {\sc ii}                &0.93$\pm$0.09  &  0.93 & 0.93\\
4711 [Ar {\sc iv}]              &2.05$\pm$0.29  & 2.76 & 2.09\\
4740 [Ar {\sc iv}]              &1.70$\pm$0.09  & 2.07 & 1.56\\
4861 H$\beta$                   &100.00$\pm$1.51& 100 & 100\\
5007 [O {\sc iii}]              &663.85$\pm$9.54&663.46 & 663.48\\
%5518 [Cl {\sc iii}]             &0.53$\pm$0.05  &0.52 & \\
5755 [N {\sc ii}]               &0.48$\pm$0.13  &  0.15 & 0.14\\
5876 He {\sc i}                 &13.76$\pm$0.23 & 13.26 & 13.14\\
6300 [O {\sc i}]                &2.58$\pm$0.06  & 1.32 & 1.79\\
6312 [S {\sc iii}]              &1.19$\pm$0.04  &1.05 & 0.83\\
6563 H$\alpha$                  &279.49$\pm$4.36&281.13 & 283.11\\
6583 [N {\sc ii}]               &6.08$\pm$0.10  & 6.11 & 6.07\\
6716 [S {\sc ii}]               &5.67$\pm$0.12  &5.68 & 6.54\\
6731 [S {\sc ii}]               &4.41$\pm$0.08  &4.41 & 5.19\\
7136 [Ar {\sc iii}]             &4.13$\pm$0.12  &  2.52 & 2.13\\
7320 [O {\sc ii}] +              &4.18$\pm$0.10  & 3.03 & 2.90\\
9069 [S {\sc iii}]              &8.10$\pm$0.20  &7.88 & 6.86\\ 
$F_{\rm corr}$(H$\beta$)$^a$               &446             & 442 & 464\\
\hline \\
%\hline
\multicolumn{4}{l}{$^a$ in units 10$^{-16}$ erg s$^{-1}$ cm$^{-2}$.}\\ 
\end{tabular}
}
\end{table}

%%%%%%%%%%%%%%%%%%%%%%%%%%%%%%%%%%%%%%%%%%%%%%%%%%%%%%%%%%%%%%%%%%%%%%%%%%%%%%%%%%%%%%

The physical conditions, and the ionic and total heavy element 
abundances in the H~{\sc ii} regions of Tol 2240--384 were derived 
following \citet{I06a} (Table \ref{tab2}). In particular for the 
O$^{2+}$, Ne$^{2+}$, and Ar$^{3+}$, we adopt
the temperature $T_e$(O~{\sc iii}) directly derived from the 
[O~{\sc iii}] $\lambda$4363/($\lambda$4959 + $\lambda$5007)
emission-line ratio. The electron temperatures $T_e$(O~{\sc ii}) and
$T_e$(S {\sc iii}) were derived from the empirical relations by
\citet{I06a}.
We used $T_e$(O~{\sc ii}) for the calculation of
O$^{+}$,  N$^{+}$, S$^{+}$, and Fe$^{2+}$ abundances and $T_e$(S {\sc iii})
for the calculation of S$^{2+}$, Cl$^{2+}$, and Ar$^{2+}$ abundances.
The electron number densities  $N_e$(O~{\sc ii}) and 
$N_e$(S~{\sc ii}) were obtained from the [O~{\sc ii}] 
$\lambda$3726/$\lambda$3729 and [S~{\sc ii}] 
$\lambda$6717/$\lambda$6731 emission-line ratios, respectively.
The low-density limit holds for the H~{\sc ii} regions
that exhibit the narrow line components considered here. 
The element abundances then do not depend sensitively on $N_e$.
The electron temperatures $T_e$(O~{\sc iii}), 
$T_e$(O~{\sc ii}), and $T_e$(S {\sc iii}), 
electron number densities $N_e$(O~{\sc ii}) and $N_e$(S~{\sc ii}), 
the ionisation correction factors ($ICF$s), and
the ionic and total O, N, Ne, S, Cl, Ar, and Fe abundances derived from the
forbidden emission lines are given in Table \ref{tab2}. It can be seen that 
the element abundances derived for the blueshifted and redshifted
components, and from the total fluxes in the UVES spectrum are 
very similar.
They are also consistent with the element abundances derived from the 
low-resolution FORS1 spectrum. However, the element abundances derived from
the medium-resolution FORS1 spectrum are somewhat different. This is
apparently due to the larger effect of the atmospheric refraction 
in the FORS1 medium-resolution spectrum resulting in a lower electron 
temperature $T_e$(O~{\sc iii}) and thus higher oxygen abundance.
For the oxygen abundance, we adopt the value 12+logO/H = 7.85$\pm$0.01. This
value is consistent within the errors with the value of 7.77$\pm$0.08 
obtained by \citet{K06}. However, for its absolute $B$ magnitude of
--21 mag, Tol 2240--384 is 
$\Delta$(12+logO/H) $\sim$ 0.7 dex below the 
oxygen abundance derived from the metallicity-luminosity relation for ELGs
by \citet{G09} (their Fig. 9).
This deviation is most likely an indication of the extreme current 
star formation in Tol 2240--384. This is similar to the lower-metallicity 
BCD SBS 0335--052E with extreme star formation, 
which for its absolute magnitude of $\sim$ -- 17 mag
is also by $\Delta$(12+logO/H) $\sim$ 0.7 dex below the value from
the relation by \citet{G09}.
The oxygen abundance in Tol 2240--384
is within the range of the oxygen abundances obtained by \citet{I07} and
\citet{IT08} for the four low-metallicity AGN candidates.
The abundance ratios N/O, Ne/O, S/O, Cl/O, Ar/O, and Fe/O 
obtained for Tol 2240--384 from the UVES spectrum and
for the four galaxies agree well.

\subsection{Photoionisation model}

We computed a photoionisation model of the narrow-line region 
to see whether the derived oxygen 
abundance is compatible with the observed temperature for a {\it bona fide} 
H~{\sc ii} region. No information is available about the morphology of the 
nebular gas, so the model from this point of view is poorly constrained. 
For the ionising source, we adopt the radiation from a starburst model computed
with STARBURST99 \citep{L99,S02} at appropriate metallicity and adopt an age of
1\,Myr (the results would not be fundamentally different for another age). The 
luminosity is adjusted to reproduce the observed H$\beta$ flux. The 
corresponding total mass of the burst is $\simeq 10^{7}M_\odot$, 
so the effects 
of statistical fluctuations to represent the ionising radiation field are 
completely negligible. We used the photoionisation code PHOTO \citep{S05}, and 
varied the elemental abundances and density distribution as free 
parameters. As already obvious from previous work \citep{SI03}, the 
He {\sc ii} $\lambda$4686 line in many H~{\sc ii} galaxies can only be
explained by an additional 
ionising source. Whether this is a population of binary stars, 
hot white dwarfs, or something else is unclear at the moment.
As in \citet{SI03}, we simply mimicked this additional X-ray component by 
bremstrahlung at  $10^6$K with the 
luminosity needed to explain the luminosity of the  He {\sc ii} $\lambda$4686 
line. This additional component has no detectable effect on the other lines.  
A  uniform density or constant pressure model did not allow us to fit all the 
constraints satisfactorily, and we had to resort to a two-density model, with 
an inner zone of density 10 cm$^{-3}$, and an outer thick shell of 
density 200 cm$^{-3}$. With this geometry, we were able to find a model, model 
M1, that reproduces the observed line ratios satisfactorily. 
In Table \ref{tab3}, 
the line ratios (in units 100$\times$$I$($\lambda$)/$I$(H$\beta$))
predicted by this model are compared to the observed narrow 
ones in the total UVES spectrum (Table \ref{tab1}). 

The chemical composition of model M1
is given in Table \ref{tab4} and compared to both the one derived from the 
empirical method and the solar value. Since we were able to reproduce the 
observed  [O~{\sc iii}] $\lambda$4363/$\lambda$5007 ratio, the abundances of 
N, O, and Ne are of 
course similar in the two approaches (note that we did not reproduce the 
[N {\sc ii}] $\lambda$5755/$\lambda$6584 ratio, but it was not used either 
in the empirical approach 
as it relies on an extremely weak line). The abundances of S, Ar, 
and Fe are less certain because the ionisation structure of these elements is 
not very well known.
We note that we had to adopt a far lower C/O ratio, than in the Sun 
\citep{A09} or in low-metallicity emission-line 
galaxies \citep[e.g., ][]{G97,IT99} 
to diminish the cooling and match the observed 
[O~{\sc iii}] $\lambda$4363/$\lambda$5007 ratio. 
This procedure is often used when no ultraviolet data are available to 
directly constrain the carbon abundance, but the carbon abundance obtained in 
this way may not be correct. Models with extra heating or more 
complex morphology and a different C/O would be equally valid from a 
photoionisation point of view. In summary, the photoionisation analysis faces 
problems that are similar to those of many {\it bona fide} low metallicity 
H~{\sc ii} regions, but does not indicate any additional problem. In terms of 
its abundance pattern, Tol~2240--384 is well within the trends exhibited
in general by metal-poor galaxies \citep{I06a}. We note only
that iron has not been depleted much, compared to H~{\sc ii} galaxies of 
similar metallicities.
Photoionisation models for Tol~2240--384 are discussed further in Sect. 6.

%%%%%%%%%%%%%%%%%%%%%%%%%%%%%%%%%%%%%%%%%%%%%%%%%%%%%%%%%%%%%%%%%%%%%%%%%%%%
%   Table 4 (abundances)
%%%%%%%%%%%%%%%%%%%%%%%%%%%%%%%%%%%%%%%%%%%%%%%%%%%%%%%%%%%%%%%%%%%%%%%%%%%%
\begin{table*} [t]
\caption{Comparison of abundances in Tol 2240--384 with the solar values} 
\label{tab4}
\centering{
\begin{tabular}{lrrrccrc} \hline \hline
%      &  Tol 2240--384    & Tol 2240--384 &Tol 2240--384    &Tol 2240--384    & Sun  \\ 
 &\multicolumn{6}{c}{Tol 2240--384}& \\ \cline{2-7}      
Element &\multicolumn{1}{c}{emp.$^{a,b}$}&\multicolumn{1}{c}{ M1$^a$} &\multicolumn{1}{c}{ M2$^a$}&\multicolumn{1}{c}{emp.$^{b,c}$}&\multicolumn{1}{c}{M1$^c$}&\multicolumn{1}{c}{M2$^c$}& Sun$^{c,d}$\\ \hline
%&(ppM)&(ppM)&(O units)&(O units)& (O units) \\ \hline
%He		&				&	.11	 & & &  \\
C  		&\multicolumn{1}{r}{...}&	3.00  & 60.00 & \multicolumn{1}{c}{...} & 0.038 & 0.65 & 0.55	\\
N  		&	3.74		&	3.75  & 4.27 & 0.052 & 0.047 & 0.046 & 0.14	\\
O  		&	71.43		&	79.00	& 92.80 &1  & 1 & 1& 1\\
Ne  	&	12.33			&	14.20	&  17.7 & 0.17 & 0.18 & 0.19 & 0.17 \\
S  		&		1.34		&	2.20  & 2.2 & 0.019 & 0.028 & 0.024 & 0.027	\\
%Cl  	&	0.03		&	0.18  & 0.0004 & 0.002 & 0.0006	\\
Ar  	&	0.30			& 0.30 & 0.30	 & 0.004& 0.004 & 0.003 &  0.005	\\
Fe  	&	3.88			&	1.75  & 2.03 & 0.054  &0.02 & 0.022 & 0.064	\\
 \hline\\
\multicolumn{8}{l}{$^a$ In units 10$^{-6}$.} \\
\multicolumn{8}{l}{$^b$ Empirical abundances.} \\
\multicolumn{8}{l}{$^c$ Relative to the oxygen abundance.} \\ 
\multicolumn{8}{l}{$^d$ Solar abundances are from \citet{A09}.} 
\end{tabular}

}
\end{table*}
%%%%%%%%%%%%%%%%%%%%%%%%%%%%%%%%%%%%%%%%%%%%%%%%%%%%%%%%%%%%%%%%%%%%%%%%%%%%

%%%%%%%%%%%%%%%%%%%%%%%%%%%%%%%%%%%%%%%%%%%%%%%%%%%%%%%%%%%%%%%%%%%%%%%%%%%%%%%%%%%%%%
%    Table 5
%%%%%%%%%%%%%%%%%%%%%%%%%%%%%%%%%%%%%%%%%%%%%%%%%%%%%%%%%%%%%%%%%%%%%%%%%%%%%%%%%%%%%%
\begin{table*}[t]
\caption{Narrow components of strong emission lines in the UVES spectrum\label{tab5}}
\centering{
\begin{tabular}{lrcrcc} \hline\hline
Line                    &\multicolumn{1}{c}{flux(blue)$^a$}&\multicolumn{1}{c}{FWHM(blue)$^b$}
&\multicolumn{1}{c}{flux(red)$^a$}&\multicolumn{1}{c}{FWHM(red)$^b$}&\multicolumn{1}{c}{separation$^c$} \\ \hline %\hline
3726 [O {\sc ii}]       &  57.3$\pm$1.5&  76.1$\pm$ 1.5&  17.3$\pm$1.7&  80.1$\pm$ 7.5&80.3$\pm$3.3 \\
3729 [O {\sc ii}]       &  62.7$\pm$0.9&  73.3$\pm$ 0.9&  22.2$\pm$0.9&  75.9$\pm$ 3.1&71.6$\pm$1.6 \\
3868 [Ne {\sc iii}]     &  68.1$\pm$0.6&  79.0$\pm$ 0.7&  31.9$\pm$0.6&  79.4$\pm$ 0.8&76.8$\pm$0.9 \\
4101 H$\delta$          &  37.4$\pm$0.7&  83.6$\pm$ 1.4&  17.1$\pm$0.7&  84.3$\pm$ 3.3&78.0$\pm$1.7 \\
4340 H$\gamma$          &  68.9$\pm$0.4&  84.1$\pm$ 0.6&  30.8$\pm$0.4&  84.4$\pm$ 1.2&77.9$\pm$0.7 \\
4363 [O {\sc iii}]      &  21.2$\pm$1.2&  82.4$\pm$ 3.6&   8.9$\pm$1.0&  71.8$\pm$ 5.5&78.8$\pm$3.7 \\
4861 H$\beta$           & 157.3$\pm$0.2&  85.2$\pm$ 0.2&  65.1$\pm$0.3&  85.8$\pm$ 0.5&78.1$\pm$0.2 \\
4959 [O {\sc iii}]      & 347.6$\pm$0.6&  79.6$\pm$ 0.2& 153.9$\pm$0.5&  79.4$\pm$ 0.4&76.6$\pm$0.2 \\
5007 [O {\sc iii}]      &1045.0$\pm$0.7&  79.9$\pm$ 0.1& 466.2$\pm$0.6&  79.9$\pm$ 0.1&76.8$\pm$0.1 \\
5876 He {\sc i}         &  25.2$\pm$0.4&  91.9$\pm$ 1.2&  10.0$\pm$0.5&  91.6$\pm$ 3.4&80.5$\pm$1.8 \\
6563 H$\alpha$          & 562.8$\pm$0.6&  86.7$\pm$ 0.1& 215.5$\pm$0.4&  87.1$\pm$ 0.2&78.5$\pm$0.1 \\
\hline \\
\multicolumn{6}{l}{$^a$Observed flux in units of 10$^{-16}$ erg s$^{-1}$ cm$^{-2}$.} \\
\multicolumn{6}{l}{$^b$In km s$^{-1}$.} \\
\multicolumn{6}{l}{$^c$Separation between blue and red components in km s$^{-1}$.} \\
\end{tabular}
}
\end{table*}
%%%%%%%%%%%%%%%%%%%%%%%%%%%%%%%%%%%%%%%%%%%%%%%%%%%%%%%%%%%%%%%%%%%%%%%%%%%%%%%%%%%%%%

%%%%%%%%%%%%%%%%%%%%%%%%%%%%%%%%%%%%%%%%%%%%%%%%%%%%%%%%%%%%%%%%%%%%%%%%%%%%%%%%%%%%%%
%   Table 6
%%%%%%%%%%%%%%%%%%%%%%%%%%%%%%%%%%%%%%%%%%%%%%%%%%%%%%%%%%%%%%%%%%%%%%%%%%%%%%%%%%%%%%
\begin{table*}[t]
\caption{Broad components of strong emission lines\label{tab6}}
\centering{
\begin{tabular}{lrrcrr} \hline\hline
    & \multicolumn{2}{c}{UVES}&&\multicolumn{2}{c}{FORS (medium)} \\ \cline{2-3} \cline{5-6}
Line                    &\multicolumn{1}{c}{flux$^a$}&\multicolumn{1}{c}{FWHM$^b$}&&
                         \multicolumn{1}{c}{flux$^a$}&\multicolumn{1}{c}{FWHM$^b$} \\ \hline %\hline
4340 H$\gamma$          &  ...~~~~~~   & ...~~~~~~  &&   1.2$\pm$0.1&2500$\pm$200\\
4861 H$\beta$           &  13.3$\pm$3.0& 380$\pm$100&&   5.0$\pm$0.1& 864$\pm$19 \\
                        &  17.2$\pm$3.0&1050$\pm$300&&   8.0$\pm$0.1&3993$\pm$93 \\
6563 H$\alpha$          &  94.7$\pm$0.4& 709$\pm$  5&&  47.3$\pm$0.2& 590$\pm$ 5 \\
                        & 169.1$\pm$0.4&2320$\pm$ 23&&  87.2$\pm$0.3&2619$\pm$37 \\
\hline \\
\multicolumn{6}{l}{$^a$Observed flux in units 10$^{-16}$ erg s$^{-1}$ cm$^{-2}$.} \\
\multicolumn{6}{l}{$^b$In km s$^{-1}$.} \\
\end{tabular}
}
\end{table*}
%%%%%%%%%%%%%%%%%%%%%%%%%%%%%%%%%%%%%%%%%%%%%%%%%%%%%%%%%%%%%%%%%%%%%%%%%%%%%%%%%%%%%%

\section{Kinematics of the ionised gas from the narrow emission lines}

We show in Table \ref{tab5} the parameters of the blueshifted and redshifted 
narrow components of the strongest emission lines in the UVES spectrum.
The two components are separated by $\sim$ 78 km s$^{-1}$
that varies only slightly from one emission line to another. 
On the other hand, the FWHMs of different lines differ somewhat. 
We note that the strong forbidden nebular emission lines 
[O~{\sc iii}] $\lambda$4959, $\lambda$5007,
[O~{\sc ii}] $\lambda$3726, $\lambda$3729, and [Ne~{\sc iii}] $\lambda$3868
have FWHMs in the range 73 -- 80 km s$^{-1}$ and are narrower than the
permitted hydrogen and helium lines. The FWHMs of the blueshifted and 
redshifted components of the weaker auroral 
[O~{\sc iii}] $\lambda$4363 emission line are 
more uncertain. The FWHMs of hydrogen lines are larger,
at $\sim$85 km s$^{-1}$. The largest FWHM of 
$\sim$92 km s$^{-1}$ is found for the He {\sc i} $\lambda$5876 emission line. 

How can we reconcile the FWHM differences of the forbidden and permitted lines?
Forbidden and permitted lines probe different parts of the
emitting regions \citep[e.g., ][]{F85}. 
It is probable that the detected emission of hydrogen and helium lines
includes a significant fraction from dense parts of
emitting regions with number densities above the critical density
of $\sim$ 10$^5$ -- 10$^7$ cm$^{-3}$ for forbidden nebular lines. At these
densities, the Balmer decrement is steeper than the pure recombination 
value because of
the contribution of the collisional excitation. The denser regions appear to be
characterised by a higher velocity dispersion. If this were the case then
we would expect to measure smaller FWHMs from H$\alpha$ to H$\delta$ because of
the decreasing fraction of emission caused by collisional excitation.
The inspection of Table \ref{tab5} shows that this is indeed the case. The widths
of the blueshited and redshifted He {\sc i} $\lambda$5876 emission lines are 
even larger than that of the narrow hydrogen lines due to the significant 
contribution of collisional excitation from 
the populated high-lying metastable level 2$^3$S. Similar evidence 
of narrow components in the permitted emission lines has been
found in some Seyfert 1 galaxies \citep[e.g., ][]{F84,M08}.

\section{Broad emission}

Using Gaussian fitting, we reassembled the H$\alpha$ and H$\beta$ emission 
lines, including broad components,
in the UVES spectrum, and the H$\alpha$, H$\beta$, and H$\gamma$ emission lines
in the medium-resolution FORS1 spectrum. 
Results of the fitting for the
H$\alpha$ emission line in the UVES spectrum are shown in Fig. \ref{fig7} and 
the parameters of the broad components for the H$\gamma$, H$\beta$, 
and H$\alpha$ emission lines are shown in Table \ref{tab6}. 
It can be seen from this Table that
the broad emission of H$\alpha$ and H$\beta$ lines could be 
fitted by two Gaussians. We note that the fitting is more uncertain for 
the H$\beta$ emission line because
of its significantly lower flux compared to that of the H$\alpha$ emission
line, especially in the UVES spectrum. The broad emission of the H$\gamma$ 
emission line in the FORS1 spectrum is yet weaker than that
of the H$\beta$ emission line and could be fitted using only a single Gaussian.
In addition, this emission is contaminated by the [O~{\sc iii}] $\lambda$4363
emission line, making the flux of the broad H$\gamma$ component more uncertain.
We also note that the fluxes of the
emission lines are lower in the FORS1 spectrum presumably due to the
smaller aperture. The FWHMs of the broad H$\alpha$ emission line in the
UVES and FORS1 spectra are in fair agreement, which is indicative of rapid
gas motion with velocities of $\ga$ 2000 km s$^{-1}$.

The observed H$\alpha$-to-H$\beta$ flux ratios of $\ga$ 10 for the broadest
components in both the UVES and FORS1 spectra are significantly higher than 
the recombination value of $\sim$ 3 expected for the low-density ionised gas
(Table \ref{tab6}). This large ratio may in part be caused by dust 
extinction.
However, the correction for extinction with $C$(H$\beta$) = 0.28 and 0.83,
respectively,
derived from the decrement of the narrow Balmer hydrogen lines in the UVES and 
FORS1 spectra (Table \ref{tab1}) implies an H$\alpha$-to-H$\beta$ flux ratio 
of $\sim$ 6. We were unable to derive the dust
extinction in the region with broad hydrogen emission. However, we could argue
that this extinction is not significantly higher than that in the region of
the narrow line emission. Otherwise, the extinction-corrected broad
H$\alpha$-to-H$\beta$ flux ratio would imply a relatively low ionised gas
density and thus the presence of a broad component in the strongest forbidden
emission line, [O~{\sc iii}] $\lambda$5007. However, this broad 
[O~{\sc iii}] emission is
not seen implying an electron number density $\ga$ 10$^{6}$ cm$^{-3}$, 
comparable to or higher than the critical electron number density for the
[O~{\sc iii}] $\lambda$4959, 5007 emission lines. The broad component
is probably present in the [O~{\sc iii}] $\lambda$4363 emission line of 
the medium-resolution FORS1 spectrum. The critical density for this auroral
line is $\sim$ 10$^8$ cm$^{-3}$. The line may therefore originate in the dense
regions, while the nebular [O~{\sc iii}] $\lambda$4959, 5007 emission lines 
do not. However, 
the [O~{\sc iii}] $\lambda$4363 emission line is much weaker and is too
close to the stronger H$\gamma$ $\lambda$4340 emission line to draw more
definite conclusions about the presence of its broad component.

In Fig. \ref{fig8}, we show
the CLOUDY model predictions of the theoretical H$\alpha$-to-H$\beta$ flux
ratio as a function of number density for three values of the ionisation
parameter log $U$ = --1, --2, and --3, respectively. The higher ionisation
parameter corresponds to stronger gas heating. For a fixed oxygen
abundance, this corresponds to a higher ionised-gas temperature.
In the CLOUDY modelling, we chose 12+logO/H=7.6.
We also adopted a power-law distribution $f_\nu$ $\propto$ $\nu^\alpha$ 
for the ionising radiation with $\alpha$ = --1 and an upper cutoff
of $\nu$ corresponding to the photon energy of 10 Ryd.

The H$\alpha$-to-H$\beta$ flux ratio at low $N_e$ in Fig. \ref{fig8} is 
constant and corresponds to the recombination value. With increasing $N_e$, the
contribution of the collisional excitation becomes important resulting in an 
increase in the H$\alpha$-to-H$\beta$ flux ratio, and the effect is stronger 
for the models with higher log $U$ where high flux ratios are achieved at lower
electron number densities because of the higher electron temperatures.

To correct the observed broad emission for extinction, we adopt for the 
extinction 
coefficients $C$(H$\beta$) the respective values obtained from the observed 
decrement of the narrow Balmer hydrogen lines (Table \ref{tab1}). Thus, 
$C$(H$\beta$) is equal to 0.28 and 0.83 for the UVES and FORS1 spectra, 
respectively. We indicate in Fig. \ref{fig8} by dash-dotted and dashed 
horizontal 
lines the extinction-corrected broad H$\alpha$-to-H$\beta$ flux ratios 
of 6.44 and 5.86 for the UVES and FORS1 spectra, respectively. These values are
significantly higher than the recombination ones and imply a high density of 
the region with broad emission. In particular, we obtain from the modelled 
H$\alpha$-to-H$\beta$ flux ratios with log $U$ = --2 the range of the electron
number densities between the two dotted vertical lines in Fig. \ref{fig8} 
of 5$\times$10$^{6}$ -- 5$\times$10$^{8}$ cm$^{-3}$ to account for the observed
ratios.
 
At a distance $D$ = 310 Mpc to Tol 2240--384, we obtain the 
extinction-corrected H$\alpha$
luminosity $L_{br}$(H$\alpha$) = 3$\times$10$^{41}$ erg s$^{-1}$ from
the UVES data. This high luminosity can probably only be explained by
the broad emission originating in a type IIn SN or an AGN, as discussed
by \citet{I07} and \citet{IT08}.
However, broad emission was present over a period of 
$\sim$ 7 years as demonstrated by the FORS1 observations in 2002 and the UVES 
observations in 2009. This rules out the hypothesis that 
the broad line fluxes are caused by type IIn SN because their H$\alpha$ fluxes
should have decreased significantly over this time interval. 

There remains the AGN scenario. Tol 2240--384  was detected
in neither the NVSS radio catalogue nor the ROSAT catalogue, 
suggesting that it is a faint radio and X-ray source,
similar to the objects discussed by \citet{IT08}. 
What about its optical spectra? Can accretion discs around
black holes in these low-metallicity dwarf galaxies account for their
spectral properties?  The spectrum of Tol 2240--384, which is similar 
to those of the four 
objects discussed by \citet{IT08}, does not show
clear evidence of an intense source of hard
nonthermal radiation: the [Ne {\sc v}] $\lambda$3426, [O~{\sc ii}]
$\lambda$3727, He {\sc ii} $\lambda$4686, [O {\sc i}] $\lambda$6300,
[N {\sc ii}] $\lambda$6583, and [S~{\sc ii}] $\lambda\lambda$6717,
6731 emission lines, which are usually found in the spectra of AGNs,
are weak or not detected.  However, if, as argued above, the density of the 
broad line region were  5$\times$10$^{6}$ -- 5$\times$10$^{8}$ cm$^{-3}$, the 
forbidden lines would be very weak or suppressed, 
except perhaps for [O~{\sc iii}] $\lambda$4363. The flux of the broad 
He {\sc ii} $\lambda$4686 line depends on both the spectral 
energy distribution of the non-thermal radiation and the ionisation 
parameter, but it is not expected to be higher than 20\% of 
the H$\beta$ line flux as seen in
\citet{S84}. A broad feature with such a low flux would be undetected
in our spectra. Some radiation from the central engine may escape to large 
distances and give rise to narrow lines. This possibility 
is roughly accounted for by the X-ray radiation included in model M1 to 
explain the observed He {\sc ii} $\lambda$4686 flux. Model M2 provides 
another solution, where the radiation field added to the stellar radiation is 
more typical of an active nucleus. We chose the broken power-law 
distribution of \citet{K00} and adjusted the luminosity of the radiation 
leaking out of the broad line region to reproduce the observed 
He {\sc ii} $\lambda$4686 flux. Under this condition, the fraction
of the H$\beta$ emission produced by the power-law ionising radiation is
$\sim$2\% of that produced by the stellar ionising radiation.
The radiation field being different from that
in model M1, some small adjustments are needed to the density distribution 
to reproduce the observed  
[O~{\sc iii}] $\lambda$5007/[O~{\sc ii}] $\lambda$3727 flux ratio, as well as 
to the abundances. Since this radiation field is more efficient at 
heating the gas, it is no longer necessary to reduce the carbon abundance to 
reduce the amount of cooling. 
In this model, the temperature in the low excitation zone is lower, therefore 
a slightly higher oxygen abundance is needed to reproduce the fluxes of 
the oxygen lines. As can be seen in Table  \ref{tab4}, the C/O abundance 
ratio in this model is much closer to both the solar value
and the value expected for the Tol 2240--384 metallicity
\citep{G97,IT99}, thus is far more 
satisfactory. This model has however some drawbacks with respect to model M1, 
the major one being that it predicts a [Ne {\sc v}] $\lambda$3426 flux 
of 5\% of H$\beta$, which should have been noted in the observed spectrum. 
To improve on photoionisation modelling by including a proper 
treatment of the broad line zone, more complete observational constraints 
would be useful.

Assuming that an AGN mechanism is responsible for the broad hydrogen
emission, we now estimate the mass of the central black hole. 
\citet{G07} derived the following relation between the central black 
hole mass and broad H$\alpha$ emission line characteristics, using the
\citet{G05} relation between the AGN continuum and H$\alpha$ luminosity
and the \citet{B06} relation between the AGN radius and continuum luminosity

\begin{equation}
\frac{M_{\rm BH}}{M_\odot}=3.0 \times 10^6 \left(\frac{L_{{\rm H}\alpha}}{10^{42} {\rm ergs\ s^{-1}}}
\right)^{0.45} \left(\frac{{\rm FWHM}_{{\rm H}\alpha}}{10^3{\rm km\ s^{-1}}}
\right)^{2.06} \label{eq1},
\end{equation}
where $L_{{\rm H}\alpha}$ is the broad H$\alpha$  luminosity, and
FWHM$_{{\rm H}\alpha}$ is the full width at half maximum of the
H$\alpha$ emission line.
The properties of the AGN in Tol 2240--384 (if it is present there)
differ from those in galaxies considered by  \citet{G07}. Therefore, the
relation in Eq. \ref{eq1} 
may not be valid for the AGN in Tol 2240--384.
In any case, we assume that this relation is valid here, since no other 
possibilities exist.
Then the mass of the black hole in Tol 2240--384 amounts to 
$M_{\rm BH}$ = 9.9$\times$10$^6$ $M_\odot$ and is higher
than the range of $M_{\rm BH}$ $\sim$ 5$\times$10$^5$ $M_\odot$ -- 
3$\times$10$^6$ derived by \citet{IT08} for a sample of four objects
and higher than the mean black hole 
mass of 1.3 $\times$ 10$^6$ $M_\odot$ found by \citet{G07} 
for their sample of low-mass black holes. 

\section{Conclusions}

We have presented 8.2m Very Large Telescope (VLT) observations with the UVES 
and the FORS1 spectrographs, and 3.5m ESO New Technology Telescope 
(NTT) $U,B,R$ imaging of the low-metallicity emission-line galaxy 
Tol 2240--384. We have studied the morphology of Tol 2240--384, the kinematics 
of the ionised gas, the element abundances, and the broad 
hydrogen emission in this galaxy. We have arrived at the following 
conclusions:

1. Image deconvolution reveals two high-surface brightness regions 
in Tol~2240--384 separated by 2.4 kpc and differing in their luminosity 
by a factor of $\sim$10. The brightest southwestern region is surrounded by
intense ionised gas emission, which strongly affects the observed $B-R$ colour
on a spatial scale of $\sim$5 kpc. This high-excitation H~{\sc ii} region is 
associated with broad H$\alpha$ and H$\beta$ emission.
Surface photometry does not indicate, in agreement with the results of image 
deconvolution and unsharp masking, the presence of a bulge in Tol~2240--384.

2. We derived the oxygen abundance 12+logO/H = 7.85$\pm$0.01 in Tol 2240--384,
which is consistent within the errors with the value of 7.77$\pm$0.08 derived
earlier by \citet{K06}.

3. The emission line profiles in the high resolution UVES spectrum reveal
the presence of two narrow components with a radial velocity difference 
of $\sim$ 78 km s$^{-1}$. Furthermore, the full widths at half maximum
(FWHMs) of the narrow lines differ. Strong forbidden nebular
lines [O~{\sc iii}] $\lambda$4959, 5007, [O~{\sc ii}] $\lambda$3726, 3729,
and [Ne~{\sc iii}] $\lambda$3868 have FWHMs of 73 -- 80 km s$^{-1}$. The
FWHMs of hydrogen lines are larger, $\sim$ 85 km s$^{-1}$ and 
decrease from H$\alpha$ to H$\delta$ emission lines. The largest
FWHM of $\sim$ 92 km s$^{-1}$ is found for the He {\sc i} $\lambda$5876
emission line. This data suggest that narrow permitted hydrogen and helium 
lines probe the denser inner parts of the emitting regions compared to the 
forbidden lines.

4. Both UVES and FORS1 spectra reveal the presence of very broad hydrogen
lines with FWHMs greater than 2000 km s$^{-1}$. 
The steep Balmer decrement of the broad hydrogen lines and 
the very high luminosity of the broad H$\alpha$ line 3$\times$10$^{41}$ 
erg s$^{-1}$
suggest that the broad emission arises from very dense and high luminosity 
regions such as those associated with supernovae of type IIn or with 
accretion discs around black holes. However, the presence of the broad 
H$\alpha$ emission over a period of 7 years rules out the SN mechanism.
Thus, the emission of broad hydrogen lines in Tol 2240--384 
is most likely associated with an accretion disc around a black hole.

5. There is no obvious spectroscopic evidence  
of a source of non-thermal hard ionising radiation in Tol 2240--384.
However, none is expected if, as we argue, the density of the broad line 
region is 5$\times$10$^{6}$ -- 5$\times$10$^{8}$ cm$^{-3}$.

6. Assuming that the broad emission in Tol 2240--384 is powered by an AGN,
we have estimated a mass for the central black hole of $M_{\rm BH}$
$\sim$ 10$^7$ $M_\odot$.

\acknowledgements
Y.I.I., N.G.G. and K.J.F. are grateful to the staff of the Max Planck 
Institute for Radioastronomy for their warm hospitality and 
acknowledge support through DFG 
grant No. FR 325/58-1. P.P. has been supported by a 
Ciencia 2008 contract, funded by FCT/MCTES (Portugal) and POPH/FSE (EC),
and by the Wenner-Gren Foundation. This research has made use of the 
NASA/IPAC Extragalactic Database (NED) which is operated by the Jet Propulsion 
Laboratory, California Institute of Technology, under contract with the 
National Aeronautics and Space Administration.
The Sloan Digital Sky Survey (SDSS) is a joint project of The University 
of Chicago, Fermilab, the Institute for Advanced Study, 
the Japan Participation 
Group, The Johns Hopkins University, the Los Alamos National Laboratory, the 
Max-Planck-Institute for Astronomy (MPIA), the Max-Planck-Institute for 
Astrophysics (MPA), New Mexico State University, University of Pittsburgh, 
Princeton University, the United States Naval Observatory, and the 
University of Washington. Funding for the project has been provided by 
the Alfred P. Sloan Foundation, the Participating Institutions, the 
National Aeronautics and Space Administration, the National Science 
Foundation, the U.S. Department of Energy, the Japanese Monbukagakusho, 
and the Max Planck Society. 

%\clearpage

%\input{ref.tex}

%\renewcommand{\baselinestretch}{1.0}

\end{document}